\documentclass[12pt]{article}
\usepackage[letterpaper,margin=1in]{geometry} 
\usepackage[utf8]{inputenc}
\usepackage{setspace} 
\usepackage{titlesec} 
\usepackage{titling} 
\usepackage{lipsum} 
\usepackage{caption} 
\usepackage{hyperref} 
\usepackage{indentfirst} 
\usepackage{authblk}

\usepackage{amsmath,amssymb}
\usepackage[style=apa, backend=biber, maxcitenames=2, mincitenames=1]{biblatex}
\usepackage[mathlines]{lineno} 

\usepackage{amsfonts}
\usepackage{graphicx}
\usepackage{color}
\usepackage{subcaption}
\usepackage{xspace}
\usepackage{xifthen}
\usepackage{multicol}
\usepackage{mathtools}
\usepackage{algorithm,algorithmicx}
\usepackage{bbm}
\usepackage{wasysym}
\usepackage[makeroom]{cancel}
\usepackage{amsmath}
\usepackage[english]{babel}
\usepackage{graphicx}
\usepackage{fontenc}
\usepackage{amssymb}
\usepackage{dsfont}
\usepackage{amsthm}
\usepackage{placeins}
\usepackage{xcolor}
\usepackage{enumerate}
\usepackage{tabularx}
\usepackage{algcompatible}
\usepackage{hyperref}
\usepackage{csquotes}
\usepackage{ mathrsfs }
\newtheorem{proposition}{Proposition}

\newtheorem{definition}{Definition}
\usepackage{comment}
\usepackage{nomencl}

\newcommand{\Dspace}{\mathcal{X}}

\newcommand{\normx}[1]{\lVert #1 \lVert}

\newcommand{\Cbin}[1]{\C{#1}{\mathrm{bin}}}

\newcommand{\nbin}{p_{\mathrm{bin}}}
\newcommand{\nbbin}{n_{\mathrm{bin}}}
\newcommand{\car}{\mathrm{card}}
\newcommand{\ellbin}{\ell_{\mathrm{bin}}}

\newcommand{\rtoc}{\textit{FindC}}
\newcommand{\ctor}{\textit{FindR}}
\newcommand{\perturb}{\textit{Perturb}}
\newcommand{\bestR}{\Rbold^{\star}(\Cbold)}
\newcommand{\bestRj}[1]{R^{\star}_{#1}(C_{#1})}

\newcommand{\bestCj}[1]{C^{\star}_{#1}(\Rbold,J,n,N)}
\newcommand{\bestC}{\Cbold^{\star}(\Rbold,J,n,N)}
\newcommand{\R}[2]{R_{#1}^{#2}}
\newcommand{\C}[2]{C_{#1}^{#2}}
\newcommand{\rdir}[1]{\boldsymbol{\gamma}_{#1}}

\newcommand{\eqdef}{:=}
\newcommand{\weight}{p}
\newcommand{\errquant}[1]{\mathcal{E}_{#1}}
\newcommand{\errglob}[1]{\epsilon_{#1}}
\newcommand{\jcal}{\mathcal{J}}
\newcommand{\Rbold}{\boldsymbol{R}}
\newcommand{\Cbold}{\boldsymbol{C}}
\newcommand{\parti}{\mathcal{P}}
\newcommand{\Ybf}{\mathbf{Y}}
\newcommand{\ybf}{\mathbf{y}}
\newcommand{\Xbf}{\mathbf{X}}
\newcommand{\setC}{\mathcal{C}}
\newcommand{\xbf}{\mathbf{x}}

\newcommand{\etabf}{\boldsymbol{\eta}}

\newcommand{\Rtrue}{\R{J_{\mathrm{true}}}{\mathrm{true}}}
\newcommand{\risk}{q^Y_{95}}
\newcommand{\riskbis}{q^Y_{99.9}}
\newcommand{\riskq}{q^Q_{90}}
\newcommand{\TSA}{\text{T-}R^2_{\text{HSIC},\omega}}
\newcommand{\Qmax}{Q_{\text{max}}}
\newcommand{\prob}{\mathbb{P}}
\newcommand{\Ks}{K_\text{s}}
\newcommand{\of}{\textit{of}}
\newcommand{\er}{\textit{er}}

\usepackage{statmath}

\def\measurehat#1{%
   \setbox0=\vbox{$\hat{#1}\hfil\break$\null\par
      \setbox0=\lastbox\unskip\unpenalty\global\setbox1=\lastbox}%
   \setbox0=\hbox{\unhbox1 \unskip\unpenalty\unskip \global\setbox2=\lastbox}%
   \setbox0=\vbox{\unvbox2 \setbox0=\lastbox}%
}
\def\doublehat#1{%
   \measurehat{#1}\dimen0=\wd0 \measurehat{\kern0pt#1}%
   \raise.35ex\rlap{\kern\dimexpr\dimen0-\wd0$\hat{\phantom{#1}}$}{\hat#1}%
}

\titleformat{\subsubsection}
{\normalfont\itshape} 
{\thesubsubsection.}
{1em}
{}
[]

\makenomenclature

\begin{document}

\def\spacingset#1{\renewcommand{\baselinestretch}%
{#1}\small\normalsize} \spacingset{1}

\date{}


\title{\bfseries Augmented Quantization: Mixture Models for Risk-Oriented Sensitivity Analysis}

\author[1]{Charlie Sire}
\author[2]{Didier Rullière}
\author[2]{Rodolphe Le Riche}
\author[3]{Jérémy Rohmer}
\author[4]{Yann Richet}
\author[4]{Lucie Pheulpin}

\affil[1]{\small Centre for Geosciences and Geoengineering, Mines Paris, PSL University}
\affil[2]{\small Mines Saint-Etienne, Univ. Clermont Auvergne, CNRS, UMR 6158 LIMOS}
\affil[3]{\small BRGM}
\affil[4]{\small ASNR}

\maketitle

\noindent
\begin{abstract}
A central question in risk analysis 
is to identify the factors that drive the system toward a specific hazardous outcome, such as the exceedance of a given threshold. When relying on numerical simulators, we propose to study the distribution of the inputs, transformed into uniform variables via their cumulative distributions, conditionally on the occurrence of the hazardous event. 
To represent this multivariate conditional distribution for sensitivity analysis, we introduce an original quantization approach based on estimating a mixture of Dirac and local uniform distributions. 
For each marginal of this mixture, a Dirac component indicates a strong influence of the corresponding variable, whereas a uniform component with wide support reflects weak influence. A notable advantage of this method is its ability to identify the regions of the input space that most strongly influence the occurrence of the risk event, while also capturing the joint effects of multiple variables. However, learning mixture models typically relies on likelihood-based methods, which are not well suited to mixtures involving singular or Dirac components. To address this, we propose an \emph{Augmented Quantization} method, a reformulation of the classical quantization problem based on the $p$-Wasserstein distance, which can be computed in very general distribution spaces. The performance of Augmented Quantization in estimating such mixture models is first demonstrated on analytical toy problems, and then applied to sensitivity analysis of both an analytical function and a practical flooding case study on a section of the Loire River. 
\end{abstract}

\vspace{1em}

\newpage
\section{Introduction}
\label{intro}

Many physical phenomena encountered in engineering, environmental science, and other applied fields are represented by complex numerical simulators. These computational models are designed to reproduce the behavior of real systems and are widely used to support decision-making processes, for example in safety assessment, risk analysis, or the optimization of operational strategies. However, these simulators typically rely on a large set of input parameters that are subject to uncertainty. 
Such uncertainty can stem from epistemic sources (\cite{helton1996treatment}), reflecting a lack of precise knowledge, as it occurs because of measurement limitations, incomplete data, or imperfect models. It can also be aleatoric in nature, arising from intrinsic randomness in the system itself, as when the inputs describe natural phenomena that often exhibit some stochasticity. To account for these different sources of uncertainty, input parameters are commonly represented as random variables characterized by known probability distributions. Given this context, sensitivity analysis (SA) provides a framework to quantify how the variability of model outputs can be attributed to the uncertainty in each input parameter (\cite{saltelli2002sensitivity, daveiga, iooss2015review}).

Among the various objectives of sensitivity analysis, this study focuses on factor mapping (see \cite{pianosi2015simple}), which seeks to identify the combinations of input variables that drive the model toward specific behaviors of interest in the output space. 
More precisely, the analysis focuses on a selected region of the model output space and seeks to identify the input factors that contribute most to the occurrence of outcomes within this region, as explored through the \emph{Target Sensitivity Analysis} (TSA) introduced in~\cite{marrel2021statistical}. 
This objective is of particular interest in risk analysis where target situations correspond to the output exceeding a given threshold, for example a pollutant concentration limit in soil pollution or the height of a dyke in flood risk assessment. 
It is worth noting that, in our framework, the target region is specified at the outset, in contrast to studies that seek the most sensitive event compatible with given input variations (\cite{roux2025maximizing}). 
The originality of our approach lies in offering a representation richer than simple numerical indices, enabling the identification of joint effects among multiple inputs that lead to the various scenarios explaining the failure event.

Let us first consider the simplified setting with one real-valued continuous random input $X$, having as cumulative distribution function (cdf) $F_X(.)$, and a random output vector $\Ybf$. 
To study the sensitivity to the input $X$ of some event $\Ybf \in \setC$, we look at the conditional distribution of 
\begin{equation*}\label{eq:FXconditionnel}
F_X(X) \mid \Ybf \in \setC
\end{equation*}
where $\setC$ denotes a subdomain of the model outputs, typically the considered extreme events. 
Studying the influence of the input  $X$ on $\Ybf \in \setC$ corresponds to analyzing if, and how, that factor drives the system toward the given risk. 
When $\Ybf \in \setC$ is independent of $X$, which means that $X$ has no influence on the occurrence of the event, the distribution of $F_X(X) \mid \Ybf \in \setC$ is uniform on $[0,1]$. In extreme situations, where only a specific value of $X$ leads to $\Ybf \in \setC$, this distribution reduces to a Dirac measure.
Thus, for one-dimensional inputs, comparing the distribution of $F_X(X)\mid \Ybf \in \setC$ to the uniform distribution on $[0,1]$ provides an interpretable information about the influence of $X$ on the occurrence of $\Ybf \in \setC$, a concept that is related to the \emph{Regional Sensitivity Analysis} proposed in~\cite{spear1980eutrophication}. 
However, when dealing with a vectorial input $\Xbf = \left(X_1,\dots, X_m\right)$, more elaborate analyses are required since it is the combined influence of several variables that drives the output into the target region $\setC$, and multiple distinct input configurations may lead to this event. In general, we need to explore the multivariate data
\begin{equation}\label{eq:FXconditionnel_m}
\big(F_1(X_1), \dots, F_m(X_m)\big) \mid \Ybf \in \setC,
\end{equation}
where $F_k$ denotes the cdf of $X_k$, thereby seeking to understand the joint inputs associated with the target region $\setC.$
Because the distribution of the random vector of Equation~\eqref{eq:FXconditionnel_m} can be difficult to interprete and visualize,  
a natural way to capture the diversity of its probability measure is to approximate it using optimal quantization (\cite{Pollard,Pages}), that is, by representing it as a mixture of Dirac measures. 
However, here it is relevant to go beyond pure Dirac mixtures and approximate the distribution with more flexible mixture models (\cite{Murray}), and specifically by combining Dirac measures with uniform components. 
Such an \emph{augmented quantization} allows the representation to degenerate into a uniform distribution on $[0,1]$ along the marginals---indicating no influence when the marginals are independent---or into a Dirac distribution, reflecting extreme dependence, or into more intricate mixture patterns.

\begin{figure}
\centering
\includegraphics[width=0.5\columnwidth,trim={0 50pt 0 0}, clip]{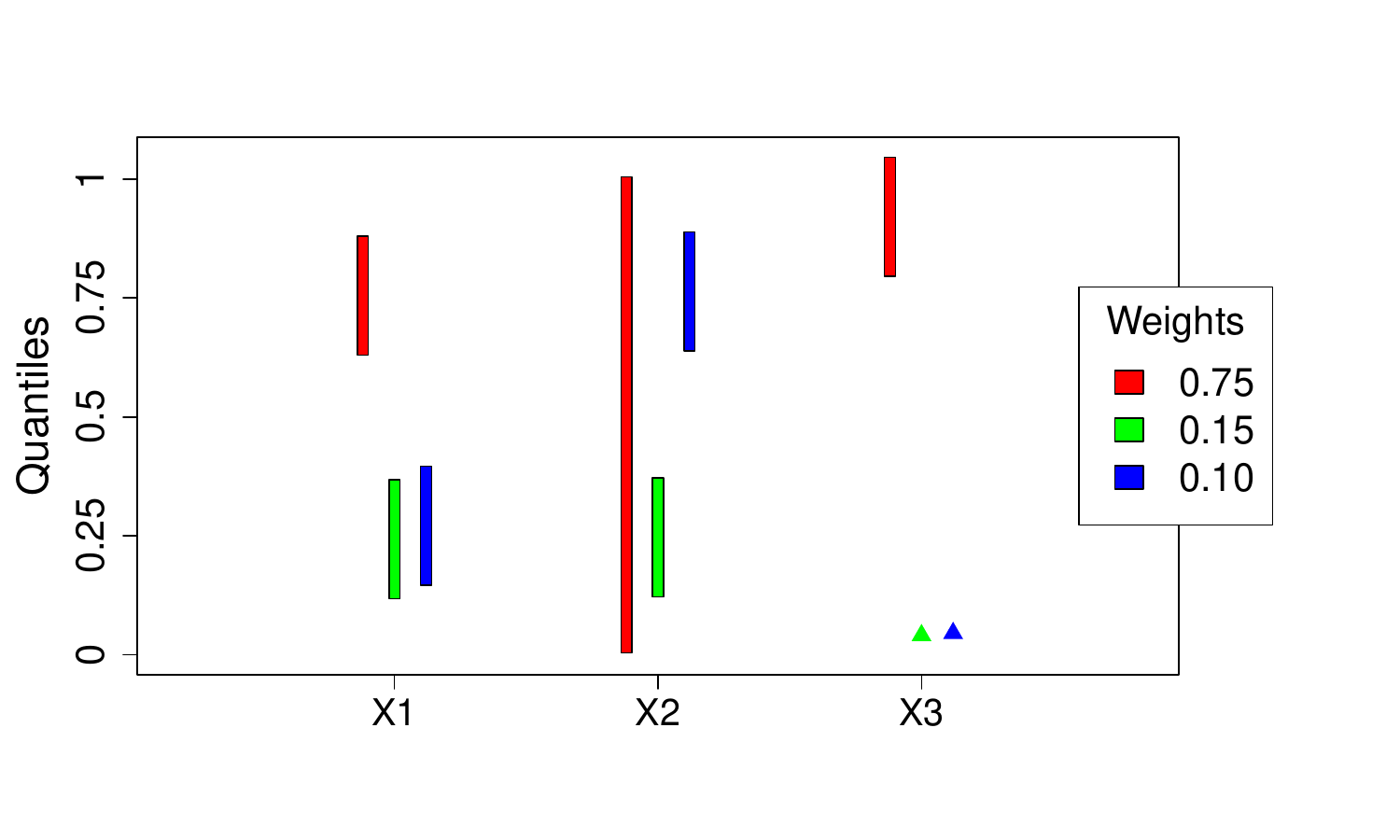}
\caption{Scenarios leading to $Y > \risk:$ approximation of
$\big(F_1(X_1), F_2(X_2), F_3(X_3)\big)\,|\, Y > \risk$ using a mixture of three distributions with Dirac and uniform independent marginals. Each distribution is associated with a color (red, green, and blue). The mixture weights are $0.75$ for the red component, $0.15$ for the green component, and $0.10$ for the blue component. A vertical bar represents a uniform distribution, while a triangle marks the location of a Dirac.}
 \label{sensi_ishigami}
\end{figure}

To motivate our study, let us briefly consider an analytical function, using a function very close to the Ishigami function\footnote{Here, we simply use $X_3^3$ instead of $X_3^4$ in the Ishigami function, so that the sign of $X_3^3$ has a joint effect with $X_1$ on the target event.} 
 (\cite{ishigami1990importance}): 
\begin{equation}\label{eq:ishigami}
Y(\Xbf) \;=\;
\sin(X_1) + a \,\sin^2(X_2) + b \, X_3^3 \sin(X_1),
\end{equation}
with $a= 4$, $b = 0.2,$ and mutually independent $X_j \sim \mathcal{U}(-\pi,\pi)$ for $1\leq j \leq 3.$ We investigate the effect of the variations of $\Xbf$ on the event $Y > \risk,$ with $\risk$ being the $95\%$-quantile of $Y$. 
Our approach approximates the conditional distribution $
\big(F_1(X_1), F_2(X_2), F_3(X_3)\big) \;\big|\; Y > \risk,$ based on $6,000$ random samples drawn from the input distribution.
The approximation relies on a mixture of three multivariate representative distributions\footnote{
Note that the number of representative distributions (three in this case) does not necessarily correspond to the number of inputs in general. See Section~\ref{appli_sensi} for a discussion of this point.}, each with independent marginals that are either uniform with different supports or Dirac, as illustrated in Figure~\ref{sensi_ishigami}. 
In simple words, any situation leading to $Y > \risk$ is close, in a specific sense that will become clear later, to one of three scenarios:
\begin{itemize}
\item In the red scenario ($75\%$ of the cases), $X_1$ lies within a small interval around its $75$th percentile (narrow uniform). $X_3$ takes values in a small interval near its maximum (narrow uniform), and $X_2$ can take any value, so it has no influence in this scenario (wide uniform).
\item In the green scenario ($15\%$ of the cases), $X_1$ falls within a small interval around its $25$th percentile (narrow uniform), $X_3$ is fixed very close to its minimum (Dirac), and $X_2$ is contained in a small interval centered at its $25$th percentile.
\item Finally, in the blue scenario ($10\%$ of the cases), $X_1$ and $X_3$ follow a configuration very similar to the green scenario, but $X_2$ lies within a small interval near its $75$th percentile.
\end{itemize}

The main configuration driving $Y > \risk$ identified here is the joint effect of $X_1$ and $X_3,$ either with $X_1$ around $\pi/2$ and $X_3$ close to its maximum (main scenario) or $X_1$ around $-\pi/2$ and $X_3$ close to its minimum. 
The effect of $X_2$ is smaller, although values of $X_2$ around $\pm \pi/2$ also encourage $Y > \risk$. 
We return to this example with additional details in Section~\ref{ishi_sec}.

This example illustrates the need for estimating mixture models combining uniform and Dirac components. 
It provides an interpretable representation of the distribution in Equation~\eqref{eq:FXconditionnel_m}, 
offering a more comprehensive understanding than numerical sensitivity indices alone do,
as it gives insights into the scenarios leading to the risk event. 
However, such models are particularly challenging to handle, mainly because most methods for estimating mixture models rely on the likelihood, which is not applicable when dealing with singular or Dirac distributions. 
This challenge drives the main objective of this paper: to establish a framework for learning mixtures of general distributions, even in settings where likelihood-based approaches cannot be used.

The paper is organized as follows. Section~\ref{overall_method} formulates the investigated problem and outlines the path leading to Augmented Quantization. Section~\ref{algo_sec} then details the implementation of the algorithm. 
Section~\ref{toy} reports the mixture models estimated on several toy problems, while Section~\ref{appli_sensi} illustrates its application to sensitivity analysis of both a toy function and a real-world simulator of flood events. Finally, Section~\ref{summary} summarizes the main results and proposes extensions to the method.

\section{Augmented quantization}
\label{overall_method}

\subsection{Challenges in mixture model estimation}

Among mixture models, Gaussian mixture models are probably the most popular, as they capture normally distributed subgroups. These mixtures are classically learned using the Expectation–Maximization (EM) algorithm (\cite{Dellaert}), which alternates between the probabilistic assignment of each observation to a cluster (E-step) through likelihood computations, and the estimation of cluster parameters by likelihood maximization (M-step). However, learning mixtures of very general distributions raises significant difficulties. The mixture representation is not always unique (see Section~\ref{prob_formu}), the number of parameters can be large, and some mixtures cannot be estimated using likelihood-based methods. In particular, when the distribution includes singular or Dirac components, the density itself may not even be defined.
Independent of likelihood considerations, our approach uses the $p$-Wasserstein distance computed between two probability measures, $\mu$ and $\nu$, on a space $\mathcal{X}$, without strong restrictions on them (\cite{villani}). 
In particular, the $p$-Wasserstein distance is defined between measures that do not have the same support or when one is discrete and the other continuous.
The $p$-Wasserstein distance is defined as follows:

$$\mathcal{W}(\mu,\nu) = \underset{\pi \in \boldsymbol{\Pi}(\mu,\nu)}{\mathrm{inf}}\left(\int_{\mathcal{X}\times\mathcal{X}}\normx{\xbf-\xbf'}^p\pi(d\xbf,d\xbf')\right)^{\frac{1}{p}},$$
where $\boldsymbol{\Pi}(\mu,\nu)$ is the set of all the joint probability measures on $\mathcal{X}\times\mathcal{X}$ whose marginals are $\mu$ and $\nu$ on the first and second factors, respectively. 
This metric represents the optimal transport cost between the two measures and appears highly relevant in our context of approximation of a set of points (the inputs for which $\Ybf \in \setC$) by a general mixture. 

As previously mentioned, finding the best mixture of $\ell$ different distributions (for a given $\ell \in \mathbb{N}$) 
generates problems with many variables. 
For instance, a mixture of $5$ Dirac measures in $4$ dimensions is described by 25 variables. 
In addition, calculating the weights and parameters of the distributions through the minimization of $p$-Wasserstein distances creates difficult, non-convex, optimization problems (\cite{merigot}). 
Clustering approaches like k-means are popular in such high-dimensional situations. 
They reduce the size of the optimization problem by decomposing it cluster by cluster. Clustering also facilitates the interpretation of the results. 
Our method, that we call \emph{Augmented Quantization} (AQ), is based on the classical k-means quantization, generalized to handle various types of distributions using the $p$-Wasserstein distance.

\subsection{Problem formulation}\label{prob_formu}

We would like to build a general mixture model that approximates the (unknown) underlying distribution of a (known) sample $(\xbf_{i})_{i=1}^{n} \in \Dspace^{n}$ with $\Dspace \subset \mathbb{R}^{m}$. 
The components of the mixture, called representatives, are probability measures on $\mathcal{X}$, that belong to a given family of probability measures denoted by $\mathcal{R}$.
More precisely, we consider $\ell \in \mathbb{N}^{\star}$ representatives named through their tag taken in $\jcal = \{1,\dots,\ell\}$.
Let $\mathcal{R}$ be a family of probability measures, the objective is approximate the distribution of $(\xbf_{i})_{i=1}^{n}$ by the mixture $R_{J}$ such that

\begin{equation}
R_{J} = \sum_{j\in \jcal} \mathbb{P}(J=j) R_{j}
\end{equation}
where $\R{j}{} \in \mathcal{R}$ for $j \in \jcal$ and $J$ is a discrete random variable independent of $(R_{1},\dots, R_{\ell})$ with weights $\weight_{j} = P(J=j)$ for $j\in \jcal$. In the example of Figure~\ref{sensi_ishigami}, $\mathcal{R}$ denotes the set of multivariate distributions with independent marginals, each being either a Dirac measure or a uniform distribution. The weights are $p_1 = 0.75,$ $p_2 = 0.15$ and $p_3 = 0.10.$

It is important to note that for a given family $\mathcal{R}$, this mixture representation is not necessarily unique, the identifiability of the problem depends on $\mathcal{R}$ (\cite{Sidney}). For instance, a mixture of the measures associated to $\mathcal{U}(0.5,1)$ and $\mathcal{U}(0,1)$ with weights $0.5$ on one hand, and a mixture of the measures associated to $\mathcal{U}(0,0.5)$ and $\mathcal{U}(0.5,1)$ with respective weights $0.25$ and $0.75$ on the other hand, are identical.

Augmented quantization, as the name indicates, generalizes the traditional quantization method and the accompanying k-means clustering. 
Let us first recall the basics of k-means. In k-means, the representatives are Dirac measures located at $\gamma$, $\mathcal{R} = \{\delta_{\gamma}, \gamma \in \mathcal{X}\}.$ For a given set of representatives $\Rbold = (\R{1}{},\dots,\R{\ell}{}) = (\delta_{\rdir{1}}, \dots, \delta_{\rdir{\ell}}) \in \mathcal{R}^{\ell},$ the associated clusters are $\Cbold = (\C{1}{},\dots, \C{\ell}{})$ with $\C{j}{} = \{\xbf \in (\xbf_{i})_{i=1}^{n}\,: \; j = \underset{i \in \jcal}{\argmin}\normx{x-\rdir{i}}\}, j\in \jcal.$ The objective is to minimize the \emph{quantization error}, 
\begin{equation}
\small
    \errquant{p}(\rdir{1},\dots,\rdir{\ell}) \eqdef \left(\frac{1}{n} \sum_{i=1}^{n} \normx{\xbf_{i} - \underset{\gamma \in 
\{\rdir{1},\dots,\rdir{\ell}\}  
}{\argmin}\normx{\xbf_{i}-\gamma} }^{p}\right)^{\frac{1}{p}}~,
\label{eq-tradi_quant_error}
\end{equation}
ensuring that each data point lies sufficiently close to at least one representative.

We now generalize the quantization error by replacing the Euclidean distance with a $p$-Wasserstein metric which, in turn, allows to rewrite the quantization error for any representatives $\Rbold$ and associated clusters $\Cbold$ that form a partition of the samples $(\xbf_{i})_{i=1}^{n}$:
\begin{equation*}
    \errquant{p}(\Cbold,\Rbold) \eqdef  \left(\sum_{j=1}^{\ell} \frac{\car\left(\C{j}{}\right)}{n} \mathcal{W}_{p}(\C{j}{}, \R{j}{})^p\right)^{\frac{1}{p}},
\end{equation*}
where $\mathcal{W}_{p}(\C{j}{}, \R{j}{})$ denotes the $p$-Wasserstein distance (\cite{Rueschendorf}) between the empirical probability measure associated with $\C{j}{}$ and the measure $R_{j}$. This notation will be used throughout the article.
 
These two quantization errors are equivalent when restricting the representatives to Dirac measures and defining the clusters as nearest neighbors to the Dirac locations $\gamma_j~,~j=1,\ldots,\ell$. We will now see how to define the Clusters $\C{j}{}~,~j=1,\ldots,\ell$ in the general case, with representatives that are not necessarily Dirac measures. 

\subsection{From k-means to Augmented Quantization}

Lloyd's algorithm is one of the most popular implementation of k-means quantization (\cite{Du}). The method can be written as described in Algorithm~\ref{alg-Lloyd}. \begin{algorithm}[H]
\textbf{Input:} $(\rdir{1},\dots,\rdir{\ell}) \in \mathcal{X}^{\ell}$~,~sample $(\xbf_{i})_{i=1}^{n}$ 
\begin{algorithmic}
\WHILE{stopping criterion not met}
\STATE \rtoc ~:~ update the clusters: $\forall j \in \mathcal J~,~ \C{j}{} = \{ \xbf \in  (\xbf_{i})_{i=1}^{n} \,: \; j = \underset{j' \in \mathcal{J}}{\argmin}\normx{\xbf-\rdir{j'}}\}$
\STATE \ctor ~:~ update the representatives: $\forall j \in \mathcal J~,~ \rdir{j} = \frac{1}{\car\left(\C{j}{}\right)}\sum_{\xbf \in \C{j}{}}\xbf
\quad,\quad  R_j = \delta_{\rdir{j}}$
\ENDWHILE
\end{algorithmic}
\textbf{Output:} Mixture $R_{J} = \sum_{j\in \jcal}\mathbb{P}(J=j) \delta_{\gamma_{j}}$
\caption{Lloyd's algorithm \label{alg-Lloyd}}
\end{algorithm} 
It is not the standard description of Lloyd's algorithm but an interpretation paving the way towards augmented quantization and the estimation of mixture distributions: 
the centroids $(\rdir{1},\dots,\rdir{\ell})$ are seen as representatives $(\R{1}{},\dots,\R{\ell}{})$ through the Dirac measures $\delta_{\rdir{}}$; 
the produced Voronoi cells are translated into a mixture distribution.

At each iteration, new clusters  $\Cbold = (\C{1}{},\dots, \C{\ell}{})$ are determined from the previously updated representatives (centroids in this case), and then new representatives $\Rbold = (\R{1}{},\dots, \R{\ell}{})$ are computed from the new clusters. These operations reduces the quantization error. We refer to these two steps as $\rtoc$ and $\ctor$. The stopping criterion is typically related to a very slight difference between the calculated representatives and those from the previous iteration.

The usual quantization is augmented by allowing representatives that belong to a predefined set of probability measures $\mathcal{R}$. 
In Augmented Quantization, $\mathcal{R}$ can still contain Dirac measures but it will typically include other measures.
Each iteration alternates between clusters $\Cbold$ during $\rtoc$ and the mixture defined by representatives and weights $(\Rbold, J)$ during $\ctor$. $\rtoc$ is now a function which partitions $(\xbf_{i})_{i=1}^{n}$ into clusters based on a set of the representatives and the random membership variable $J$.
$\ctor$ is a function which generates a set of representatives from a partition of $(\xbf_{i})_{i=1}^{n}$.
Algorithm \ref{augmented_algo} is the skeleton of the Augmented Quantization algorithm.

\begin{algorithm}[h]
\textbf{Input:} $\Rbold = (\R{1}{},\dots, \R{\ell}{}) \in \mathcal{R}^{\ell}$, samples $(\xbf_{i})_{i=1}^{n}$
\begin{algorithmic}
\STATEx $J\in \mathcal{J}$ r.v.  with $\mathbb{P}(J = j) = \frac{1}{\ell} $
\STATEx $(\R{}{\star}, \C{}{\star}, \mathcal{E}^{\star}) \gets (\emptyset, \emptyset, +\infty)$
\WHILE{stopping criterion not met}
\STATE{Update clusters: $\Cbold \gets \rtoc(\Rbold,J)$ }
\STATE{Perturb clusters: $\Cbold \gets \perturb(\Cbold)$ }
\STATE{Update mixture: $\Rbold \gets \ctor(\Cbold)$, $J$ r.v. with $\mathbb P(J=j) = \frac{\car\left(\C{j}{}\right)}{n},\: j \in \mathcal{J}$ }
\STATE{Update the best configuration: 
\IF{$\errquant{p}(\Cbold,\Rbold)<\mathcal{E}^\star$}
   $\mathcal{E}^\star \gets \mathcal{E} ~,~ \Cbold^{\star} \gets \Cbold ~,~ \Rbold^{\star} \gets \Rbold ~,~ J^{\star} \gets J $}
\ENDWHILE
\end{algorithmic}
\textbf{Output:} Optimal mixture $\R{J^{\star}}{\star}$
\caption{Augmented Quantization algorithm \label{augmented_algo}}
\end{algorithm}

Lloyd's algorithm can be seen as a descent algorithm applied to the minimization of the quantization error of Equation \eqref{eq-tradi_quant_error}. It converges to a stationary point which may not even be a local optimum (\cite{selim}). 
Although Lloyd's algorithm converges generally to stationary points that have a satisfactory quantization error, an example is provided in the Supplementary Material (\cite{charliesire_2025}) to illustrate that its generalization to continuous distributions does not lead to a quantization error sufficiently close to the global optimum.  To overcome this limitation, an additional mechanism for exploring the space of mixture parameters is needed. To this aim, we propose a perturbation of the clusters called $\perturb$, which takes place between $\rtoc$ and $\ctor$. The need for a perturbation has also been described in classical k-means methods where it takes the form of a tuning of the initial centroids when restarting the algorithm (\cite{capo}).

Before detailing these algorithm steps, some theoretical elements are required to motivate our implementation choices.

\subsection{Properties of quantization errors}\label{theory}

Here, we first define the different errors and related propositions concerning our AQ procedure. Let $\Rbold = (R_{1}, \dots, R_{\ell})$ be $\ell$ probability measures and $\Cbold = (C_{1}, \dots, C_{\ell})$ be $\ell$ disjoint clusters of points in $\mathcal{X} \subset \mathbb{R}^m$. Let us denote $n_{j} = \car\left(C_{j}\right)$ for $j\in \jcal$, and $n = \sum_{j=1}^{n}n_{j}$.

\begin{definition}[Quantization error]
The \emph{quantization error} between $\Cbold$ and $\Rbold$ is defined by
\begin{equation}\label{quanti_error}
\errquant{p}(\Cbold,\Rbold) \eqdef \left(\sum_{j}^{\ell}\frac{n_{j}}{n}\mathcal{W}_{p}(C_{j}, R_{j})^{p}\right)^{1/p}.
\end{equation}
\end{definition}

\begin{definition}[Global error]
The \emph{global error} between $\Cbold$ and $\Rbold$ is defined by
\begin{equation}\label{global_error}
    \errglob{p}(\Cbold,\Rbold) \eqdef \mathcal{W}_{p}\left(\bigcup_{j=1}^{\ell} C_{j}, R_{J}\right),
\end{equation}
where $J\in \jcal$ is a random variable such that $\mathbb P(J=j) = \frac{n_{j}}{n}.$
\end{definition}

The quantization error aggregates the local errors between the clusters and the representatives, while the global error characterizes the overall mixture. The quantization error provides a natural measure of clustering performance, and this clustering serves here as a means to decompose the minimization of the global error. This decomposition is justified by Proposition~\ref{glob_and_quanti}, which shows that a low quantization error ensures a low global error.

\begin{proposition}[Global and quantization errors]
The global error between a clustering $\Cbold$ and a set of representatives $\Rbold$ is lower than the quantization error between them: 
\begin{align*}
\errglob{p}(\Cbold,\Rbold) 
 = \mathcal{W}_{p}\left(\bigcup_{j=1}^{\ell} C_{j}, R_{J}\right)  \leq \left(\sum_{j}^{\ell}\frac{n_{j}}{n}\mathcal{W}_{p}(C_{j}, R_{j})^{p}\right)^{1/p} = \errquant{p}(\Cbold,\Rbold)~.
\end{align*}\label{glob_and_quanti}
\end{proposition}
\noindent The proof is provided in Appendix~\ref{proof_glob_quanti}. Our approach therefore focuses on the quantization error and aims to identify the clustering $\Cbold$ that minimizes the \emph{clustering error}, defined below.

\begin{definition}[Clustering error]\label{def_clust_error}
The clustering error of a partition $\Cbold$ is the quantization error between $\Cbold$ and $\bestR$, its associated optimal representatives,

\begin{equation}\label{clustering_error}
  \errquant{p}(\Cbold) \eqdef \errquant{p}(\Cbold, \bestR)  
   = \underset{\Rbold \in \mathcal{R}^{\ell}}{\mathrm{min}} \:\errquant{p}(\Cbold, \Rbold),
   \end{equation}
  
with $\bestR \eqdef \underset{\Rbold \in \mathcal{R}^{\ell}}{\argmin}\:\errquant{p}(\Cbold,\Rbold).$
\end{definition}

These optimal representatives can be obtained using Proposition~\ref{optim_rep}.

\begin{proposition}[Optimal representatives]\label{optim_rep} The optimal representatives of a clustering $\Cbold$ can be optimized independently for each cluster,
\begin{equation*}
\bestR = (\bestRj{1},\dots, \bestRj{\ell})\;\;\mathrm{ where }\;\; \bestRj{j} \eqdef \underset{r \in \mathcal{R}}{\argmin}\:\mathcal{W}_{p}(C_{j},r), \:j\in \jcal. 
\end{equation*}
\end{proposition}

This results trivially from the fact that $\errquant{p}$ is a monotonic transformation of a sum of independent components. This is one reason to treat mixture estimation as a clustering task, which reduces the dimensionality of the optimization.
The optimal representatives are then based on $p$-Wasserstein distance minimizations in the space of probability measures $\mathcal R$. This problem is further addressed in Section~\ref{ctor} when seeking representatives from given clusters.

\section{Algorithm steps}\label{algo_sec}

The steps $\rtoc$, $Perturb$ and $\ctor$ of the Augmented Quantization algorithm can now be presented in details. We will explain how they contribute to reducing the quantization error which was discussed in Section~\ref{theory}. Appendix~\ref{detailedalgo} provides a detailed implementation of the algorithm, including a discussion of implementation aspects such as the stopping criteria. The associated code is provided in the Supplementary Material (\cite{charliesire_2025}).

\subsection{Finding clusters from representatives}\label{rtoc}

At this step, a mixture distribution $R_J$ is given through its $\ell$ representatives $\Rbold = (R_{1}, \dots, R_{\ell})$ and the associated membership random variable $J$ such that $P(J=j) = \weight_{j}~,~j=1,\ldots,\ell$. 
Clustering is performed by, first, creating $N$ samples from the mixture distribution: the membership variable $j$ is sampled from $J$, and each point is drawn from $R_j$. Second, the data points are assigned to the cluster to which belongs the closest of the $N$ samples. Before returning to the algorithm, we provide a theoretical justification for it.

\vskip\baselineskip

Let $(\Xbf_{i})_{i=1}^{n} \in \mathcal{X}^{n}$ be a random sample of the above $R_J$ mixture distribution.
$(J_{i})_{i=1}^{N}$ are i.i.d. samples with the same distribution as $J$, and $(\Ybf_{i})_{i=1}^{N}$ i.i.d. samples with $\Ybf_{i} \sim R_{J_{i}},\:i=1,\dots,N.$
We define the following clustering,
\begin{equation*}
\bestC \eqdef \Bigl( \bestCj{1},\dots,\bestCj{\ell} \Bigl)\\
\end{equation*}
with  $\bestCj{j} \eqdef \{\Xbf_{k}\; s.t. \; J_{I_{N}(\Xbf_{k})} = j\; ,\; 1\leq k\leq n\}$ and $I_{N}(\xbf) \eqdef \underset{i =1,\dots,N}{\argmin}{\mid\mid \xbf-\Ybf_{i} \mid \mid }~.$
We work with general distributions that are combinations of continuous and discrete distributions, $\mathcal{R}_{s} \eqdef \{\beta_{\mathrm{c}}R_{\mathrm{c}} + \beta_{\mathrm{disc}}R_{\mathrm{disc}}, R_{\mathrm{c}} \in \mathcal{R}_{\mathrm{c}}, R_{\mathrm{disc}} \in \mathcal{R}_{\mathrm{disc}}, \beta_{\mathrm{c}} + \beta_{\mathrm{disc}} = 1\}$.

$\mathcal{R}_{\mathrm{c}}$ and $\mathcal{R}_{\mathrm{disc}}$ are the set of measures associated to almost everywhere continuous distributions with finite support, and, the set of measures associated to discrete distributions with finite support, respectively. 

In the family of probability measures $R_{s}$, the above clustering is asymptotically consistent: its clustering error (see Definition~\ref{def_clust_error}) is expected to tend to zero as $n$ and $N$ increase.

\begin{proposition}[Asymptotic clustering consistency.]\label{prop_findclust}

If $R_{j}\in \mathcal{R}_{s}$ for $j \in \jcal$ and $\Xbf_{i} \sim R_{J_{i}},\; i=1,\dots,n$ with $(J_{i})_{i=1}^{n}$ i.i.d. sample with same distribution as $J$, then 

$$\underset{n,N\rightarrow+\infty}{lim} \mathbb{E}\left(\errquant{p}(\bestC)\right) = 0.$$
\end{proposition}

The proof along with further details about $\mathcal{R}_{s}$ are given in the Supplementary Material.\\

The objective of the $\rtoc$ procedure is to associate a partition of a sample $(\Xbf_{i})_{i=1}^{n}$ from the representatives $\Rbold$ and their probabilistic weights defined by the random variable $J$. 
It is described in Algorithm~\ref{rtocalgo}.
\begin{algorithm}
\textbf{Input:} Sample $(\Xbf_{i})_{i=1}^{n}$, $\Rbold = (\R{1}{},\dots,\R{\ell}{})$, $N$, $J$ r.v. $\in \jcal$ 
\begin{algorithmic}
\STATE{$C_{j} = \emptyset,\: j\in\jcal$}
\STATE{$(j_{i})_{i=1}^{N}$ N independent realizations of $J$}
\STATE{$(\ybf_{i})_{i=1}^{N}$ N independent realizations, $\ybf_{i}$ sampled with associated measure $R_{j_{i}}$}
\FOR{$\xbf \in (\xbf_{i})_{i=1}^{n}$}
\STATE{$I(\xbf) \gets \underset{i =1,\dots,N}{\argmin}{\mid\mid \xbf-y_{i} \mid \mid }$}
\STATE{$\C{j_{I(\xbf)}}{} \gets \C{j_{I(\xbf)}}{} \cup \xbf$}
\ENDFOR
\caption{$\rtoc$}
\label{rtocalgo}
\end{algorithmic}
\textbf{Output:} Partition $\Cbold = (\C{1}{},\dots, \C{\ell}{})$ 
\end{algorithm}

The $\rtoc$ algorithm is consistent in the sense of the above Proposition~\ref{prop_findclust}.

It is important to note that this procedure is not intended to serve as a robust classifier in the Machine Learning sense. In particular, a point $\xbf$ may originate from representative $R_1$ while being associated with cluster $C_2$. The goal, however, is solely to construct a partition corresponding to the mixture measure $R_J$.

\subsection{Perturb clusters}\label{perturb}

Once clusters are associated to representatives, the $\perturb$ step is required to explore the space of partitions of $(\xbf_{i})_{i=1}^{n}$.  An illustration of why this step is important in our context is provided in the Supplementary Material, using the example of a mixture of two uniform distributions which, without perturbation, cannot be identified from an given initial clustering. Thus, a relevant cluster perturbation should be sufficiently exploratory. 
To increase convergence speed, we make it greedy by imposing a systematic decrease in quantization error through the clustering error.

\begin{proposition}[Greedy cluster perturbation]\label{prop_perturb}
~\\
Let $\Cbold = (C_{1}, \dots, C_{\ell})$ be a clustering, $G(\Cbold)$ a set of perturbations of this clustering such that $\Cbold \subset G(\Cbold)$, and $\perturb(\Cbold)\eqdef \underset{\Cbold' \in G(\Cbold)}{\argmin}\: \errquant{p}(\Cbold').$

Trivially, $$\errquant{p}\left(\perturb(\Cbold)\right) \leq \errquant{p}(\Cbold).$$

\end{proposition}
The quantization error decrease comes from the inclusion of the current clustering in the set of perturbations.
Here, we choose to perturb the clusters through the identification of the points contributing the most to the quantization error. Our cluster perturbation consists in, first, identifying the elements to move which defines the $split$ phase and, second, reassigning them to other clusters in what is the $merge$ phase. These steps are detailed in Appendix~\ref{appendix_perturb}. 

\subsection{Finding representatives from clusters}
\label{ctor}
 
Given a clustering $\Cbold = (\C{1}{},\dots, \C{\ell}{})$ obtained after peturbation, the $\ctor$ step searches for the associated representatives $\bestR = (\bestRj{1},\dots,\bestRj{\ell})$ that are optimal in the sense that 
$\bestRj{j} \eqdef  \underset{r \in \mathcal{R}}{\argmin}\mathcal{W}_{p}(\C{j}{},r)$.
In practice, $\mathcal{R}$ is a parametric family  $\{r(\etabf), \etabf \in \mathbb{R}^s\}$.
Instead of performing an optimization of $\mathcal{W}_{p}(\C{j}{},r)$ over the multidimensional distribution, the minimization is approximated by optimizing separately over each dimension, that is, over the $p$-Wasserstein distances between the marginals. 
Such an approximation is numerically efficient because the Wasserstein distance in 1D can be easily expressed analytically for two probability measures $\mu_1$ and $\mu_2$ (\cite{Panaretos}): 
\begin{equation}
    \mathcal{W}_{p}(\mu_{1}, \mu_{2}) \eqdef \left(\int_{0}^{1} \mid F_{1}^{-1}(q) - F_{2}^{-1}(q) \mid ^{p}dq\right)^{\frac{1}{p}},
\end{equation}
where $F_{1}$ and $F_2$ are the cumulative distribution functions. 
A detailed example of the $\ctor$ function with Dirac and uniform distributions are described in the Supplementary Material. In this example, the analytical minimization of the $p$-Wasserstein distance has only a single local optimum, which is inherently the solution to the problem. 
Situations with multiple local optima may happen, but the analytical expression of the distance is, in practice, a strong asset in favor of the numerical tractability of $\ctor$. It must be noted that, while optimizing the $p$-Wasserstein distance on the marginals produces representatives with independent marginals, the resulting mixture $R_J$ can still show strong dependency across marginals, which may help reveal joint effects for our sensitivity analysis. This point will be further discussed in Section~\ref{appli_sensi}. 

\section{Toy problem for mixture model estimation}\label{toy}

In this section, we evaluate the performance of the method on samples representing mixtures of uniform and Dirac distributions, as considered in our sensitivity analysis application briefly introduced in the Introduction. Specifically, we construct a $m$-dimensional mixture of distributions that have independent marginals, where each marginal is either a uniform distribution with support width $0.25$, $0.5$, $0.75$, or $1$, or a Dirac measure.

Technically, the family of representatives consists of all distributions $R$ of the form  
$R = R_{1} \times \dots \times R_{m}$  where 
each $R_{k}$ depends on parameters $(\alpha_{1},\dots,\alpha_{m}, a_{1}, \dots, a_{m}, \sigma_{1},\dots, \sigma_{m}):$
\begin{itemize}
\item If the boolean $\alpha_k = 1$, $R_{k}$ is a uniform between $a_k - \frac{\sigma_k}{2}$ and $a_k + \frac{\sigma_k}{2},$ with $\sigma_k \in \{0.25,0.5,0.75,1\}.$
\item If $\alpha_k = 0,$ $R_{k}$ is a Dirac at $a_k \in [0,1],$ and $\sigma_k = 0.$
\end{itemize}

\begin{figure*}
  \begin{subfigure}[t]{0.48\textwidth}\centering
\includegraphics[width=0.98\linewidth,trim={0 50pt 0 0}, clip]{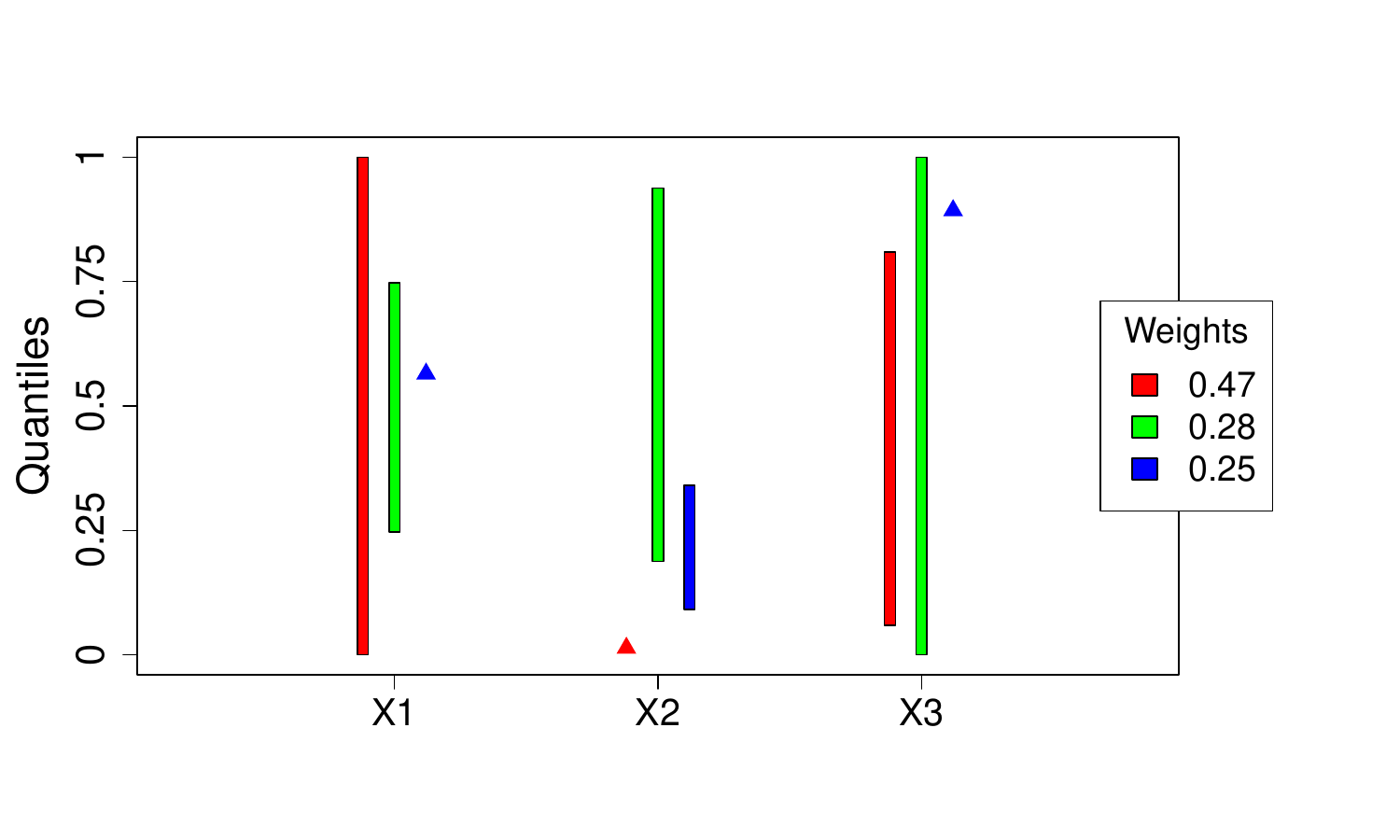}
\subcaption{True mixture.}
 \label{rtrue_hyb}
 \end{subfigure}%
 ~ 
\begin{subfigure}[t]{0.48\textwidth}\centering
\includegraphics[width=0.98\linewidth,trim={0 50pt 0 0}, clip]{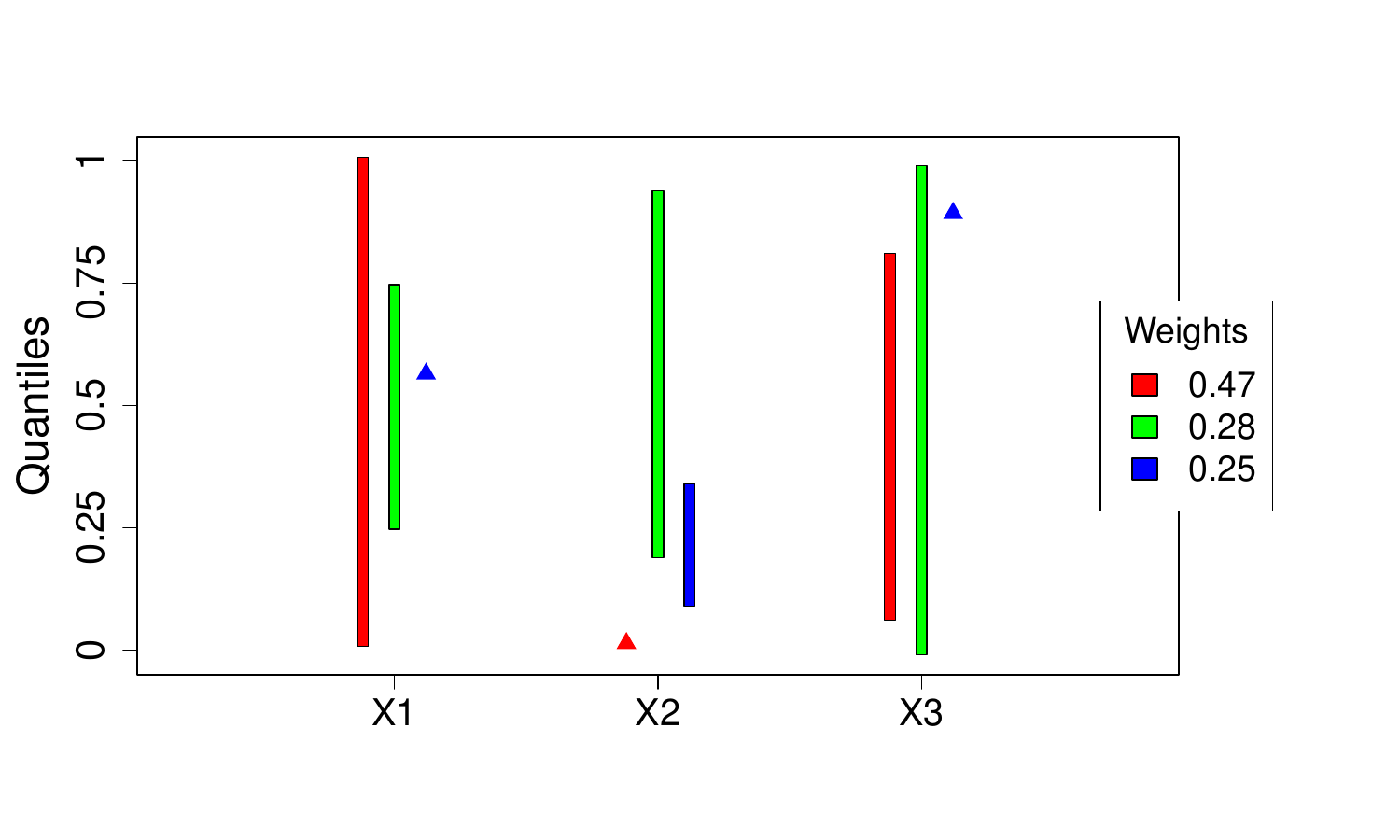}
 \subcaption{Estimated mixture with AQ.}
 \label{raq_hyb}
  \end{subfigure}
\caption{Example of the estimation of a $3$-dimensional mixture of three representatives (shown in red, green, and blue) with Dirac or uniform marginals. 
A vertical bar represents a uniform distribution, while a triangle marks the location of a Dirac. 
The mixture weights are $0.47$ for the red representative, $0.28$ for the green one, and $0.25$ for the blue one. Figure~\ref{rtrue_hyb}:~True mixture. Figure~\ref{raq_hyb}:~Estimated mixture.}
     \label{r_hyb}
\end{figure*}

The analytical solution for the $\ctor$ step with this family of representatives is detailed in the Supplementary Material. The idea is thus to study $m$-dimensional samples $\mathcal{S}_{\mathrm{sensi}} = \left(\mathbf{X}_{i}\right)_{i=1}^{n}$ 
that are associated with a true mixture $\Rtrue$ built from $\ell$ representatives of the form described above and to best represent their distribution with AQ algorithm. An illustrative example is provided in Figure~\ref{r_hyb}. Each representative is assigned a distinct color, and its corresponding weight in the mixture is indicated on the right side of the plots. Figure~\ref{r_hyb} shows that our method accurately captures the true mixture, with representatives and weights nearly identical to the true values. 

\begin{figure*}
  \begin{subfigure}[t]{0.48\textwidth}\centering
\includegraphics[width=0.98\linewidth]{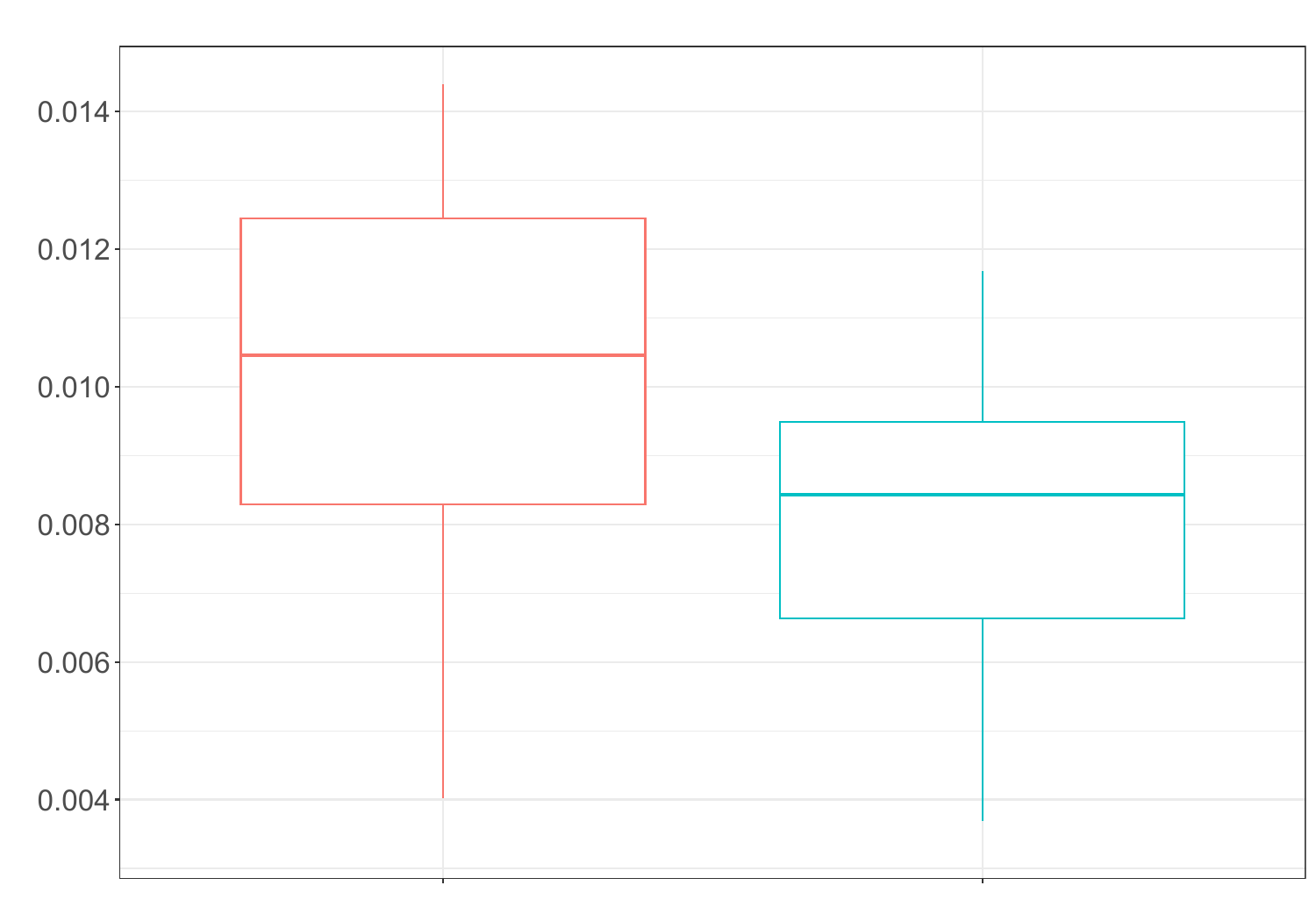}
\subcaption{Quantization error distribution of the true mixtures (red) and of the mixtures estimated by AQ (blue).}
 \end{subfigure}%
 ~ 
\begin{subfigure}[t]{0.48\textwidth}\centering
\includegraphics[width=0.98\linewidth]{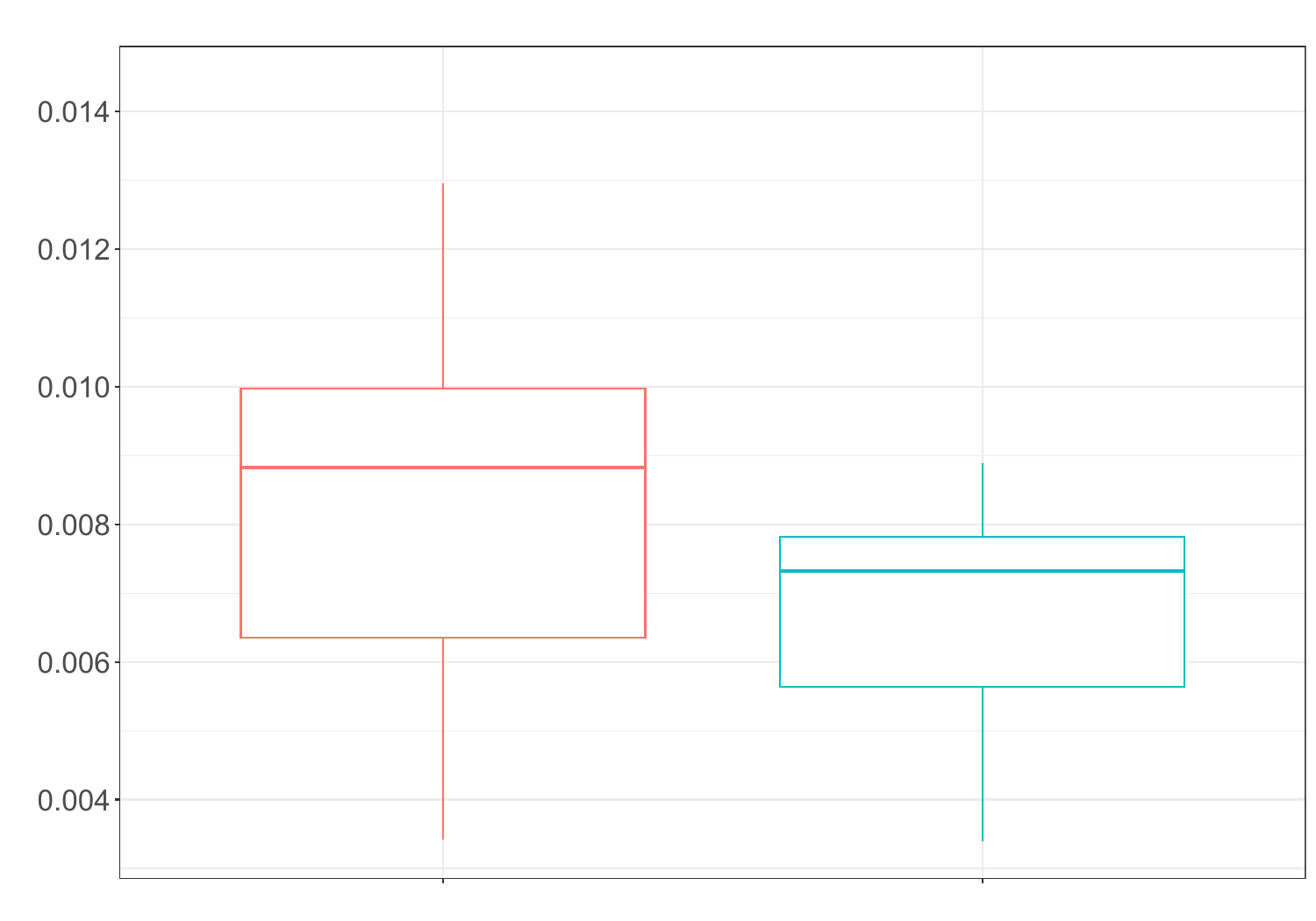}
 \subcaption{Global error distribution of the true mixtures (red) and of the mixtures estimated by AQ (blue).}
  \end{subfigure}
  \caption{Distributions of the errors in the $3-$dimensional test case of mixtures with Dirac and uniform representatives.}
     \label{boxplots_hyb}
\end{figure*}

To assess the robustness of the method, $15$ different mixtures of $200$ points are considered, 
with parameters randomly selected. 
These parameters cover 

$$(\alpha_{1},\dots,\alpha_{m}, a_{1},\dots,a_{m}, \sigma_{1},\dots,\sigma_{m}),$$
which define the representatives, and the weights of the mixtures. For each mixture, the obtained quantization error of Equation~\eqref{quanti_error} and the global error of Equation~\eqref{global_error} will be computed, to quantify the errors between the samples and the estimated mixture. The errors between the samples and the true mixtures will be calculated as well, as they quantify the sampling error and provide a basis for comparison. They are called \emph{sampling errors}. Note that the samples $\mathcal{S}_{\mathrm{sensi}}$ are obtained with quasi-Monte Carlo to work with very small sampling errors. The distributions of the errors are shown in Figure~\ref{boxplots_hyb}. The errors obtained with the AQ algorithm are highly satisfactory, as they are smaller than the sampling error, which is a promising outcome. In other words, AQ is able to identify mixtures that are closer to the sample (in terms of both quantization and global errors) than the true mixture itself.  Figure~\ref{boxplots_hyb} also highlights that the global errors are lower that the quantization errors, as stated in Proposition~\ref{glob_and_quanti}.

Results for other mixtures, such as Gaussian mixtures, are provided in the Supplementary Material to demonstrate that the AQ approach is far more general than merely estimating Dirac or uniform mixtures.

\FloatBarrier

\section{Application to sensitivity analysis}\label{appli_sensi}

The proposed general mixture model is useful to perform sensitivity analysis on computer codes with random vector inputs $\Xbf = \left(X_1,\dots,X_m\right)$ and a random output $\Ybf$. 
The objective is to estimate a mixture model for the distribution of Equation~\eqref{eq:FXconditionnel_m},
$ \big(F_1(X_1), \dots, F_m(X_m)\big) \,\big|\, \Ybf \in \setC, $
where $F_j(X_j)$ denotes the unconditional cumulative distribution function of $X_j$. 
The investigated representatives are described in Section~\ref{toy}: they are $m-$dimensional distributions with independent marginals, each marginal being a Dirac measure or a uniform distribution with support width $0.25, 0.5, 0.75$ or $1$. 

The dependencies within the distribution of $X_1, \dots, X_m$ should now be briefly discussed. 
One of the objectives is to highlight the joint effects of certain inputs on the event $\Ybf \in \setC$ by identifying dependencies in the conditional distribution of Equation~\eqref{eq:FXconditionnel_m}. 
Recall that even if the representative distributions forming the mixture have independent marginals, the resulting mixture can still effectively reveal strong dependencies among some components $F(X_k)$ conditional on $\Ybf \in \setC$, as can multivariate histograms (see~\cite{Arbenz,laverny2021dependence}). 
A mixture of Dirac measures is a case of extreme dependence in the distribution. 
Although the independence of $X_1, \dots, X_m$ is not required to capture the scenarios leading to the risk event, comparing the conditional dependencies with the initial dependence structure of the inputs provides valuable insights into whether these dependencies are intrinsic or induced by the conditioning event $\Ybf \in \setC$. 
In cases where $X_1, \dots, X_m$ are dependent with a known copula, one option is to remove the dependency by applying the AQ method to a conditional sample obtained via the Rosenblatt transformation of the inputs~\cite{darling1952test}, thereby facilitating the comparison and interpretation between the initial and the conditional dependence structures.

The analytical function similar to the Ishigami function (introduced in Section~\ref{intro}) is now revisited and followed by an application to floodings. 
In the Supplementary Material (\cite{charliesire_2025}), the G-function from \cite{sobol1998quasi} is examined with ten input variables, and representatives whose marginals are either uniform with large support or Dirac measures, illustrating the use of AQ in a screening context. 
This example is interesting because it has analytical parameters that control the influence of the input variables (see \cite{marrel2009calculations}).

\FloatBarrier
\subsection{Analytical test case}\label{ishi_sec}

Considering the function of Equation~\eqref{eq:ishigami}, we study the influence of $\Xbf$ on the occurrence of the event $Y > \risk$, with $\risk$ denoting the $95\%$-quantile of $Y$. 
The value of $\risk$ is empirically estimated as $\risk \approx 5.63$. Then, from independent samples $\left(\Xbf_i\right)_{i=1}^{6000}$, we obtain $n = 290$ realizations of 
\[
\big(F_1(X_1), F_2(X_2), F_3(X_3)\big) \;\big|\; Y > \risk,
\]
and we aim to approximate this conditional distribution using a mixture of three distributions with Dirac and uniform marginals. The results are presented in Figure~\ref{sensi_ishigami} for a mixture of three representatives. 
The brief analysis in Section~\ref{intro} highlighted the following influences:
\begin{enumerate}
\item The joint effect of $X_1$ and $X_3$ is the main driver of $Y > \risk$ and appears to always play a significant role, with
\begin{itemize}
\item $X_1$ around $\pi/2$ and $X_3$ close to $\pi$, which is the predominant configuration ($75\%$ of the cases),
\item $X_1$ around $-\pi/2$ and $X_3 \approx -\pi$ ($25\%$ of the cases).
\end{itemize}
\item The influence of $X_2$ is smaller, but values of $X_2$ near $\pi/2$ or $-\pi/2$ are identified as a contributing factor in $25\%$ of the cases.
\end{enumerate}
These insights are consistent with the different terms of the function. 
Focusing only on the individual influences, the variable $X_3$ appears to be the most influential, with two Dirac measures and a narrow uniform distribution on its marginals, indicating that the values of $X_3$ leading to $Y > \risk$ are highly concentrated. The variable $X_1$ is less influential but still impacts the target event, with $3$ narrow uniform distributions of support width $0.25$. Finally, as noted earlier, $X_2$ appears to be the least influential variable, as $75\%$ of its probabilistic mass corresponds to a uniform distributions between $0$ and $1$. 

\begin{figure}
    \centering
    \includegraphics[width=0.8\textwidth]{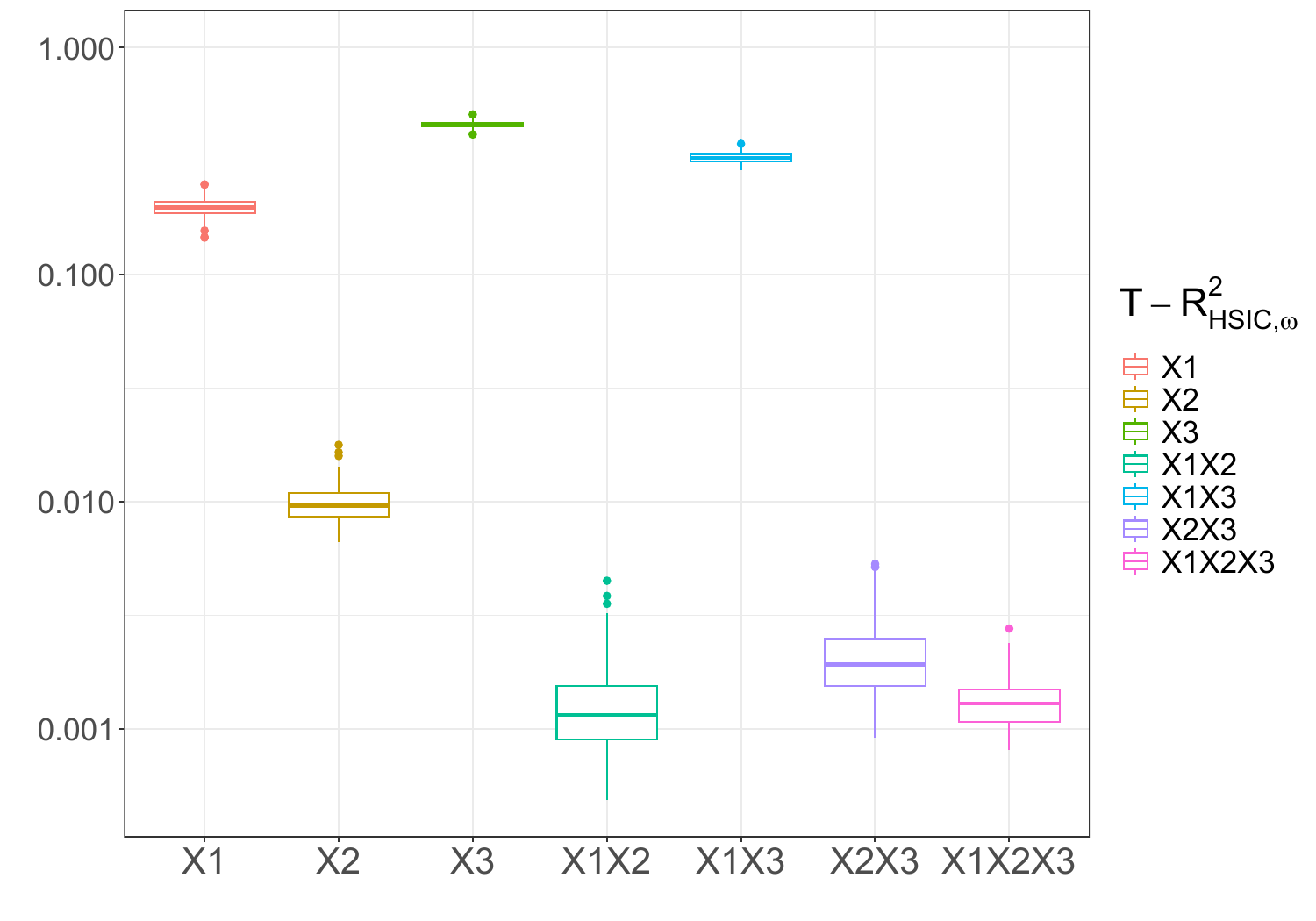}
    \caption{
HSIC $\TSA(X_i, Y)$ indices using ANOVA-like decomposition estimated from samples of size $6000$ in the analytical test case, plotted on a logarithmic scale.}
    \label{hsic_ishi}
\end{figure}

This analysis can be compared to what is obtained with the \emph{Normalized Target HSIC indices}, denoted $\TSA(X_i,Y)$, as presented in \cite{marrel2021statistical}. 
These indices are equivalent to HSIC indices between $X_i$ and a weighted transformation $\omega(Y)$ of $Y$, where $\omega(Y)$ is a smooth approximation of the indicator function $\mathds{1}_\setC$. 
HSIC indices can be defined with any tuple of inputs, using an ANOVA-like decomposition as proposed in ~\cite{da2021kernel}. 
Here, we use an exponential weighting function (see \cite{marrel2021statistical} and \cite{raguet2018target}), although 
other smoothing functions could also be employed (see \cite{spagnol}). 
The $\TSA(X_i,Y)$ indices are computed from independent samples $\left(\Xbf_i\right)_{i=1}^{6000}$, and the results are reported in Figure~\ref{hsic_ishi}. 
The HSIC indices confirm the AQ analysis: $X_3$ is the main individual driver, its joint effect with $X_1$ is also central. The impact of $X_2$ is smaller but still identified, while all the other joint effects seem negligible.

It becomes clear that, although the sensitivity indices provide a meaningful ranking of input influences, the AQ representation offers complementary insights by highlighting the different scenarios that lead to $Y > \risk$. 
In particular, it identifies the regions of the input space associated with the occurrence of the risk, either as intervals of restricted width (uniform with narrow support) or, in extreme cases, as single critical values (Dirac).

An important parameter of the method is the number $\ell$ of mixture representatives. 
When AQ is used for mixture model estimation, a common approach is to test the method with several values of $\ell$ and examine how the resulting quantization error evolves with $\ell$. A suitable value of $\ell$ can then be determined by the user using an elbow-type criterion. In the context of sensitivity analysis, to facilitate interpretation, one idea is to retain only the representatives associated with probabilistic weights larger than a minimum threshold $p_{\text{min}}$. The number of representatives $\ell$ can then be increased until at least one weight falls below this threshold. In the applications presented in this paper, we set $p_{\text{min}} = 10\%$ to focus on the main effects.

\FloatBarrier
\subsection{Flooding test case}\label{flood_sec}

\begin{figure}
    \centering
    \includegraphics[width=0.65\textwidth]{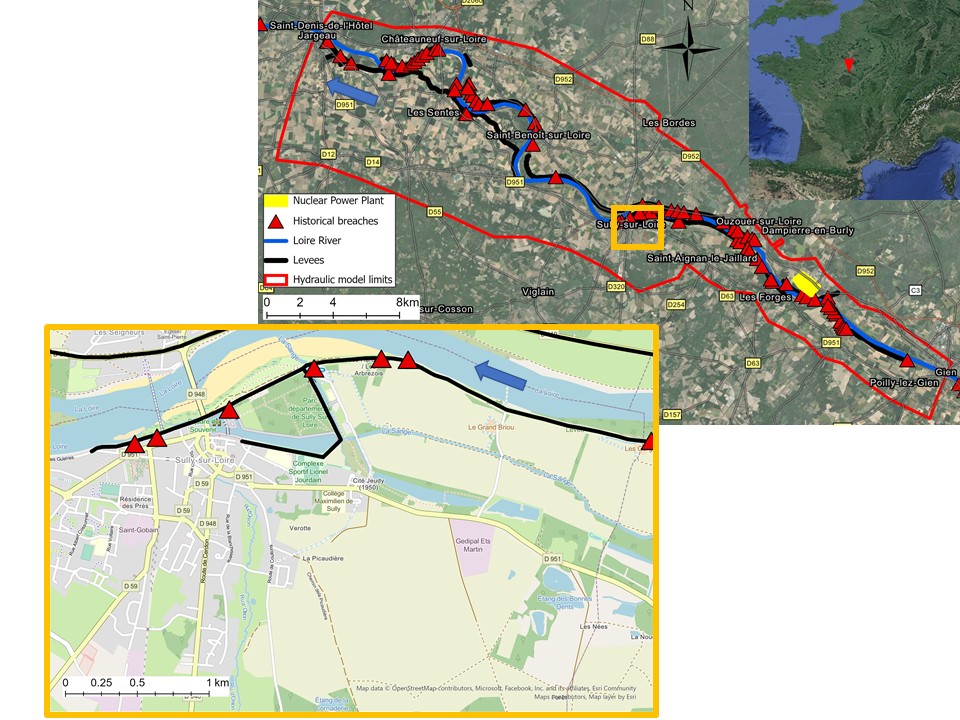}
    \caption[Study area of the flooding test case]{Study area of the flooding test case, including the levees and the historical breaches (\cite{osm}).}
    \label{fig_loire}
\end{figure}

We now investigate a computer code that generates flood maps of a specific area as output. 
The inputs correspond to environmental features that may cause flooding, and the output $Y$ is the total volume of water inundating a specific land area. 
The case study focuses on a section of the Loire River near Orléans, France, which is flanked by levees on both banks and has a history of significant flooding in the 19th century. The area studied, including historical levee breaches (\cite{Maurin}), is depicted in Figure~\ref{fig_loire}.

To simulate the river flow of the Loire River between Gien and Jargeau over a distance of 50 km, the French Authority for Nuclear Safety and Radiation Protection (ASNR) has built a hydraulic model using the open-source TELEMAC-2D simulator (\cite{Pheulpin}). 
The model incorporates an upstream hydrograph and a calibration curve as boundary conditions and has been calibrated using well-known flood events by adjusting the roughness coefficient (Strickler coefficients). 
Breaches are considered as well, leading to the study of four variables, considered as independent:
\begin{itemize}
    \item The maximum flow rate $\Qmax$ follows a Generalized Extreme Value distribution, established using the Loire daily flow rate at Gien.
    \item The hydraulic roughness coefficient $\Ks$ is a calibration parameters. This coefficient constitutes a calculation artifact and is not observed in real-world data sets. Therefore, a triangular distribution is employed and its mode corresponds to the calibration value.
\item The overflow \of\ is related to the breaching of the dyke and represents a criterion for breach initiation: if the overflow is measured at \of\ (in~m) below or above the levee crest, a breach occurs. The distribution is uniform between $-0.2$ and $0.2.$
\item The erosion rate \er\ describes the vertical extent of the breach during the simulated time, which follows a uniform distribution.
\end{itemize}
To keep the computation cost feasible, we restrict our attention to the left bank sector of the river, specifically Sully-sur-Loire. 
There, we have simulated seven breaches whose sites and lengths correspond to those of the actual history. 
As flooding is a rare event, very high quantiles on the total volume of water inundating the land are investigated.
More precisely, we are interested in the event  
\[
\normx{Y} > \riskbis,
\]  
where $\normx{Y}$ denotes the Euclidean norm of the matrix representing the flood map, and $\riskbis$ corresponds to the empirical $0.999$-quantile of $\normx{Y}$.  

Computing this quantile is not particularly challenging in practice, since it is assumed that such extreme events occur only when $\Qmax$ exceeds its $0.9$-quantile, denoted by $\riskq$. This allows us to restrict the sampling to the truncated distribution of $\Qmax$ above this threshold. Under this assumption, we can write  
\[
\prob\big(\normx{Y} > \riskbis\big)  
= \prob\big(\normx{Y} > \riskbis \,\mid\, \Qmax > \riskq\big)\,\prob\big(\Qmax > \riskq\big).  
\]  
Hence, investigating the $0.99$-quantile of $\normx{Y}$ conditional on $\Qmax > \riskq$ is equivalent to studying the unconditional $0.999$-quantile of $\normx{Y}$. 
The same truncated sampling is used to generate $500$ flood maps belonging to the investigating domain.

\begin{figure}
\centering
\includegraphics[width=0.7\columnwidth,trim={0 50pt 0 0}, clip]{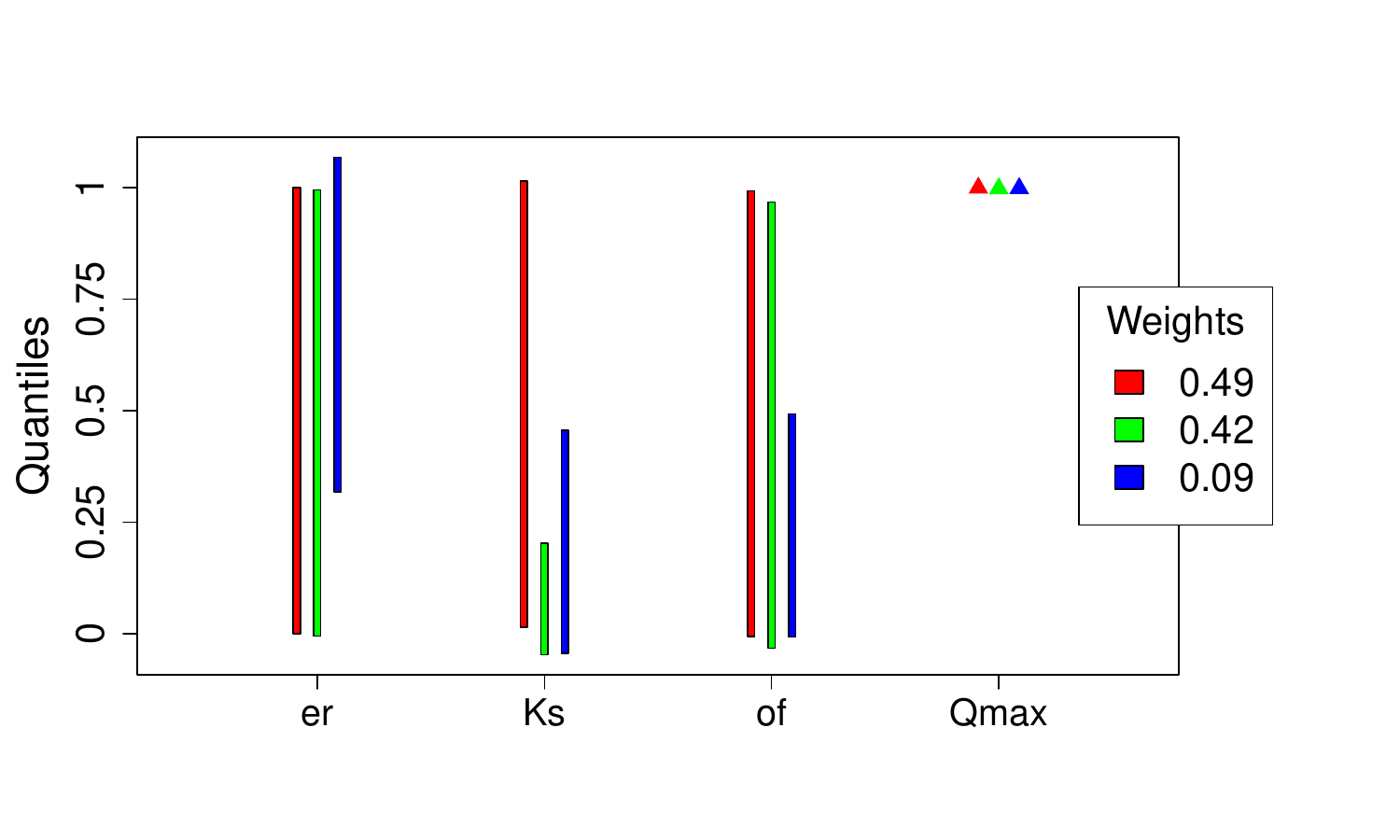}
\caption{Scenarios leading to $\normx{Y} > \riskbis$ in the flooding test case: mixture of three distributions with Dirac and uniform independent marginals. Each distribution is associated with a color (red, green, and blue). The mixture weights are $0.49$ for the red component, $0.42$ for the green component, and $0.09$ for the blue component. A vertical bar represents a uniform distribution, while a triangle marks the location of a Dirac.}

 \label{augm_flood_hyb}
\end{figure}

Figure~\ref{augm_flood_hyb} presents the obtained mixture, with the same modelling as that previously used in Section~\ref{toy}. 
As expected, the maximum flow rate, $Q_{\mathrm{max}}$, is the most influential variable, as only Dirac distributions are identified at quantiles around $0.999$, indicating that very high quantiles of $Q_{\mathrm{max}}$ are associated with such extreme events.

The impact of the roughness coefficient $\Ks$ is also evident, as lower values of $\Ks$ tend to promote more intense flooding. 
This effect is shown by the green representative, which accounts for $42\%$ of the scenarios and corresponds to $\Ks$ values close to their minimum, characterized by a narrow uniform distribution. 
The blue representative, encompassing $9\%$ of the scenarios, is associated with $\Ks$ values below its median, further emphasizing this influence. 

The influence of \er\ and \of\ is much smaller, since in $91\%$ of the scenarios (red and green representatives) these parameters can take almost any value, with wide uniform distributions identified. The limited influence of these breaching parameters can be attributed to the extremely high flow rates $Q_{\mathrm{max}}$ involved, which render their impact almost negligible, since water overtops the levees in any case. However, the blue representative, which includes the remaining cases, reveals a joint effect of \er\ and \of: the combination of relatively high erosion rates and small overflow thresholds appears to drive severe flooding. This is an expected effect, as large breaches (high \er) that are easily initiated (low \of) represent additional risk factors. 

The target HSIC indices are also computed to complement the analysis. 
However, since a very high quantile of $\lVert Y \rVert$ is being targeted, classical Monte Carlo sampling is not appropriate. 
Although using the truncated distribution $\lVert Y \rVert \mid \Qmax > \riskq$, based on specific values of $\Qmax$, is possible, it introduces a bias in the estimated effects involving $\Qmax$. 
Since it is by far the most influential variable, we do not consider it for the computation of the HSIC indices, and focus on the three remaining variables \er, $\Ks$ and \of. The development of a dedicated method to estimate TSA indices for very rare events would be of significant interest, but lies beyond the scope of the present work.

\begin{figure}
    \centering
    \includegraphics[width=0.8\textwidth]{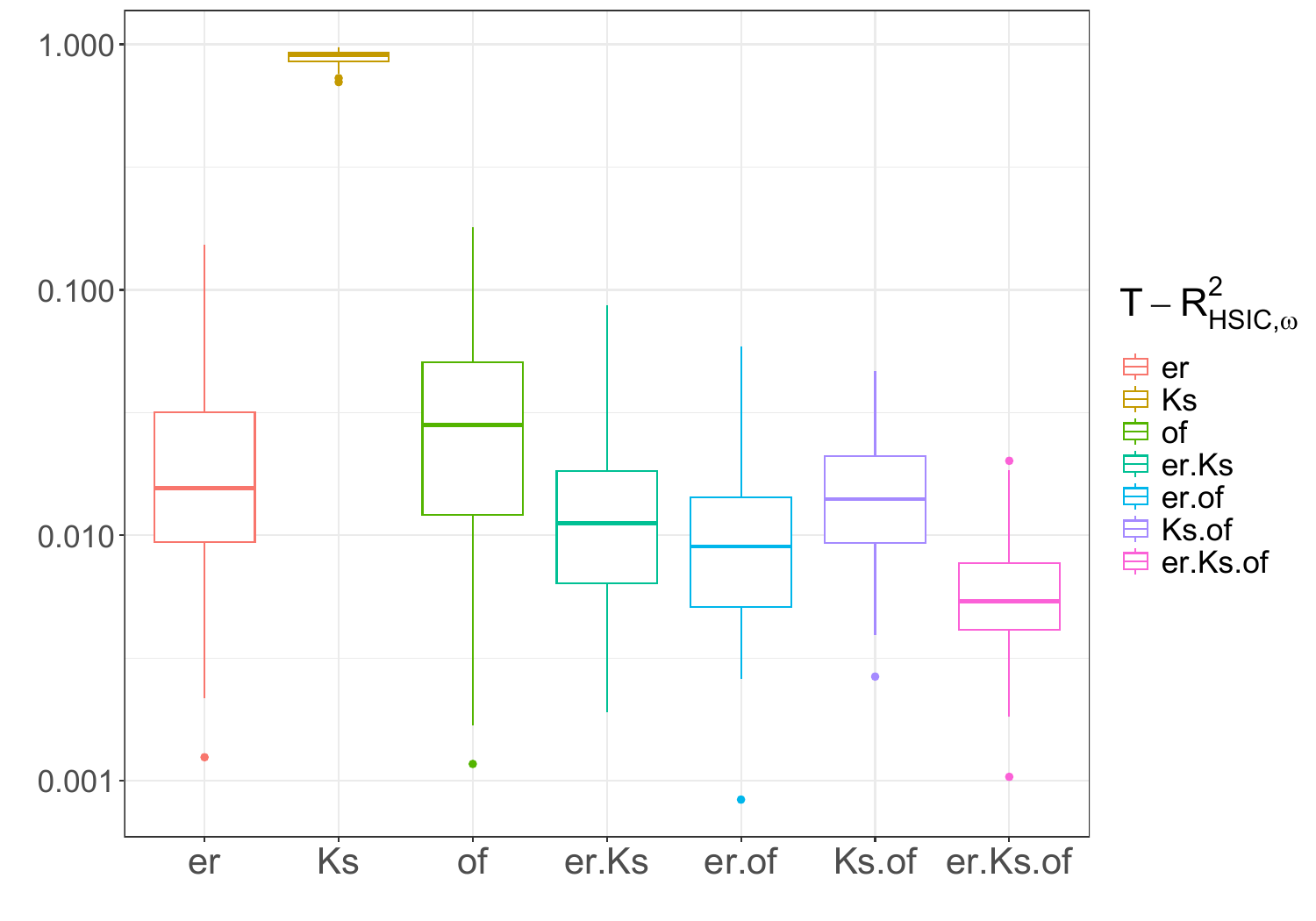}
    \caption{Target HSIC $\TSA(X_i, Y)$ estimated from samples of size $10^4$ in the flooding test case, plotted on a logarithmic scale.}
    \label{hsic_flood}
\end{figure}

The results obtained using the truncated distribution (Figure~\ref{hsic_flood}) highlights that the influence of $\Ks$ on the most severe flooding events is dominant, clearly surpassing all other effects (excluding $\Qmax$, which was not considered). The remaining factors were identified in only 9\% of the scenarios in our AQ analysis, which confirms this dominance. Although the joint effect of \er\ and \of\ was not detected by the HSIC indices, the relatively higher contribution of \of\ compared to \er\ is consistent with our AQ results, as the blue representative shows a narrower uniform distribution for overflow than for erosion rate.

\section{Summary and perspectives}\label{summary}

The present work introduces an innovative method to estimate general mixture models for target sensitivity analysis. The underlying idea is straightforward: to identify the configurations that most strongly influence a risk event, we examine the distribution of the inputs conditional on the occurrence of this event that are first transformed into uniform variables on $[0,1]$. An effective and intuitive way to gain insights into the influence of each input is then to approximate its conditional distribution using a mixture of uniforms or Dirac measures, thereby highlighting deviations from the uniform baseline on $[0,1]$. To obtain such hybrid mixtures, we propose the Augmented Quantization (AQ) algorithm, a reformulation of the classical $k$-means procedure based on the Wasserstein distance. This approach makes it possible to estimate mixture models even in cases where the likelihood is not defined. 

With AQ, the target event can be defined in a very general way. We have illustrated this with exceedances over thresholds and shown that AQ not only identifies the most influential inputs leading to the occurrence of the risk event, but also highlights the regions of the input space associated with it (input mapping). The flooding case study is particularly illustrative in this respect, as it emphasizes the dominant role of the river flow rate and the secondary influence of the breaching conditions. Beyond improving our understanding of risk conditions, this twofold analysis can support uncertainty characterization in many decision-support contexts:
(1) in scenario definition, within a \emph{scenario discovery} framework (\cite{bryant2010thinking}), and 
(2) in the development of environmental policies, by enabling a reassessment of environmental threats not only in terms of their severity, as in \cite{puy2020current}, but also in terms of their occurrence. Furthermore, by defining the target event with respect to the functioning of the underlying physical system (e.g., the goodness-of-fit between model responses and observations), AQ can contribute to model verification, as discussed by \cite{razavi2021future}. In the flooding case for instance, AQ can highlight the restricted range of the hydraulic roughness coefficient on which a complementary calibration phase should particularly focus.

While the present work focuses on sensitivity analysis, the AQ method is much more general and can be applied to estimate a broad class of mixture models including gaussian mixtures or other distributions. Then, it can also represent multivariate distributions directly, without uniform transformation or conditioning. Nevertheless, further developments remain to be investigated.\\~\

The proposed method is, for now, designed to provide mixtures of representative distributions with independent marginals, since the $\ctor$ step is decomposed into $m$ one-dimensional optimizations. As already noted, even under this assumption the resulting mixtures can still reveal compelling dependencies across marginals, which is one of the main objectives of the clustering step, so this is not an obstacle to identifying joint effects. Nevertheless, extending the method to representatives that incorporate dependence structures across marginals would be a valuable enhancement for mixture model estimation. One possible direction is to consider known parametric families of copulas and incorporate their parameters into the optimization of the $\ctor$ step.

An important parameter of the method is the number $\ell$ of representative distributions. As discussed in Section~\ref{ishi_sec}, this number is chosen to ensure that the minimum probabilistic weight of the mixture remains below a threshold, thereby facilitating interpretability. In a more general setting, cluster perturbation could be used to adapt the number of clusters as it relies in splitting several clusters and on a subsequent merge to recover the same number of clusters $\ell$ (see Appendix~\ref{appendix_perturb}). This $Perturb$ step then temporarily increases the number of clusters before reducing it again, and strong variations in the quantization error observed during this step may indicate that the number of representatives should remain higher, in the spirit of an active-learning–type adjustment.

Finally, regarding the computational cost of the current AQ algorithm, the sample size is limited to at most a few thousand observations due to the reliance on greedy procedures. The $Perturb$ operation is the most computationally demanding step, as the points contributing the most to the quantization errors are moved sequentially, and all possibilities are tested for each move (see Appendix~\ref{appendix_perturb}). To reduce this burden, it may be worthwhile to explore partial search strategies, although this may reduce precision.


\bigskip
\begin{center}
{\large\textbf SUPPLEMENTARY MATERIAL}
\end{center}

\noindent\textit{All codes related to the AQ method:} A Git repository containing supplementary technical materials and R notebooks to reproduce all experiments related to the test cases described in the article, as well as additional toy problems referenced in the main text. See \cite{charliesire_2025}.

\printbibliography

\FloatBarrier
\appendix 

\section{Proof of Proposition~\ref{glob_and_quanti}}\label{proof_glob_quanti}

Let $\boldsymbol{R} = (R_{1}, \dots, R_{\ell})$ be $\ell$ probability measures and $\boldsymbol{C} = (C_{1}, \dots, C_{\ell})$ be $\ell$ disjoint clusters of points $\in \mathcal{X} \subset \mathbb{R}^m$.
In addition, let us denote 
 \begin{itemize}
    \item $n_{j} = \car\left(C_{j}\right)$ for $j\in \jcal$, and $n = \sum_{j=1}^{\ell} n_{j}$
     \item $\mu_{C}^{j}$ the empirical measure associated to $C_{j}$
     \item $\boldsymbol{\Pi}_{j}$ as the set of all coupling probabilities $\pi_{j}$ on $\mathcal{X}\times \mathcal{X}$ such that $\int_{\mathcal{X}} \pi_{j}(\xbf,\xbf')d\xbf = R_{j}(\xbf')$ and $\int_{\mathcal{X}} \pi_{j}(\xbf,\xbf')d\xbf' = \mu_{C}^{j}(\xbf)$
     \item $\pi^{\star}_{j} = \underset{\pi_{j} \in \boldsymbol{\Pi}_{j}}{\mathrm{inf}}\int_{\mathcal{X}\times\mathcal{X}}\normx{\xbf-\xbf'}^p\pi_{j}(\xbf,\xbf')$
     \item $p_{j} = \frac{n_{j}}{n}$
     \item $\mu_{UC}$ the empirical measure associated to $\bigcup_{j=1}^{\ell} C_{j}$
     \item $\boldsymbol{\Pi}_{\mathrm{global}}$ as the set of all coupling probabilities $\pi$ on $\mathcal{X}\times \mathcal{X}$ such that $\int_{\mathcal{X}} \pi(\xbf,\xbf')d\xbf = R_{J}(\xbf')$ and $\int_{\mathcal{X}} \boldsymbol{\Pi}_{j}(\xbf,\xbf')d\xbf' = \mu_{UC}(\xbf)$
 \end{itemize}

By definition of $R_{J}$, we have $R_{J} = \sum_{j=1}^{\ell}p_{j}R_{j}$.
Similarly, we have $\mu_{UC} =  \sum_{j=1}^{\ell}p_{j}\mu_{C}^{j}$.
\begin{align*}
    \errquant{p}(\boldsymbol{C},\boldsymbol{R}) &= \left(\sum_{j}^{\ell}\frac{n_{j}}{n}\mathcal{W}_{p}(C_{j}, R_{j})^{p}\right)^{\frac{1}/{p}} \\
    &= \left(\sum_{j=1}^{\ell}\frac{n_{j}}{n}\int_{\mathcal{X}\times\mathcal{X}}\normx{\xbf-\xbf'}^p\pi^{\star}_{j}(d\xbf,d\xbf')\right)^{\frac{1}{p}} \\
    & = \left(\int_{\mathcal{X}\times\mathcal{X}}\normx{\xbf-\xbf'}^p\sum_{j=1}^{\ell} p_{j}\pi_{j}^{\star}(d\xbf,d\xbf')\right)^{\frac{1}{p}} \\
    & = \left(\int_{\mathcal{X}\times\mathcal{X}}\normx{\xbf-\xbf'}^{p}\tilde{\pi}(d\xbf,d\xbf')\right)^{\frac{1}{p}} 
    \end{align*}
with $\tilde{\pi}=\sum_{j=1}^{\ell} p_{j}\pi_{j}^{\star}$.

Since for $j\in \jcal, \pi_{j}^{\star} \in \boldsymbol{\Pi}_{j}$, 

\begin{align*}
    \int_{\mathcal{X}} \tilde{\pi}(d\xbf,\xbf') &= \sum_{j=1}^{\ell} p_{j}\int_{\mathcal{X}}\pi_{j}^{\star}(d\xbf,\xbf') \\
    &=\sum_{j=1}^{\ell} p_{j}R_{j}(\xbf')d\xbf \\
    &= R_{J}(\xbf') ~.
\end{align*}
Similarly, $\int_{\mathcal{X}} \tilde{\pi}(\xbf,\xbf')d\xbf' = \mu_{UC}(\xbf)$.
This shows that $\tilde{\pi} \in \boldsymbol{\Pi}_{\mathrm{global}}$.

Finally, 
\begin{align*}
    \errglob{p}(\boldsymbol{C},\boldsymbol{R}) &= \mathcal{W}_{p}(\bigcup_{j=1}^{\ell} C_{j}, R_{J})\\&=
\underset{\pi \in \boldsymbol{\Pi}_{\mathrm{global}}}{\mathrm{inf}}\left(\int_{\mathcal{X}\times\mathcal{X}}\normx{\xbf-\xbf'}^p\pi(d\xbf,d\xbf') \right)^{\frac{1}{p}}\\
&\leq \errquant{p}(\boldsymbol{C},\boldsymbol{R})
\hspace{4cm} \square
\end{align*}

\section{Details on the \emph{Perturb} step}\label{appendix_perturb}

This appendix presents our proposition for perturbing the clusters, which consists of two steps: first, a \emph{split} of certain clusters, followed by a \emph{merge} phase that identifies the best way to return to $\ell$ clusters.

\subsection{\textit{split} phase}

First, let us define for a given cluster $ \C{j}{},$ its associated local error:
$$w_{p}(C_{j}) \eqdef \mathcal{W}_{p}(C_{j},\bestRj{j}) = \underset{r\in \mathcal{R}}{\mathrm{min}}\: \mathcal{W}_{p}(C_{j}, r).$$
In this phase, among the clusters $(\C{1}{}, \dots, \C{\ell}{})$, the $\ellbin$ clusters with the highest local errors $w_{p}$ will be split. Here, $\ellbin$ is an integer less than $\ell$. Their indices make the $\mathrm{indexes_{bin}}$ list.
During the $split$ phase, for each cluster $\C{j}{} , j \in \mathrm{indexes_{bin}}$, a proportion of $\nbin$ points are sequentially removed and put in a sister ``bin'' cluster. 
A point $\xbf^\star$ is removed from the cluster if the clustering composed of the cluster after point removal and the bin cluster has the lowest error. 

The values of $\ellbin$ and $\nbin$ determine the magnitude of the perturbations. Their values will be discussed after the $merge$ procedure is presented. 
Algorithm~\ref{algo-split} sums up the $split$ procedure.

\begin{algorithm}
\textbf{Input:} a sample $(\xbf_{i})_{i=1}^{n}$, a partition $\Cbold = (\C{1}{},\dots, \C{\ell}{})$, $\nbin \in [0,1]$, $\mathrm{indexes_{bin}} = \{j_{1},\dots, j_{\ellbin}\}$
\begin{algorithmic}
\FOR{$j \in \mathrm{indexes_{bin}}$}
\STATE{$\Cbin{j} \gets \emptyset$}
\STATE{$\nbbin \gets \nbin \car\left(\C{j}{}\right)$}
\WHILE{$\car\left(\Cbin{j}\right) < \nbbin$}
\STATE{$\xbf^{\star} \gets \argmin_{\xbf \in \C{j}{}} \errquant{p}(\Cbold^{split}_{j}(\xbf)) 
 \quad\text{ where } \Cbold^{split}_{j}(\xbf) ~=~ (\C{j}{}\setminus \xbf,\Cbin{j}\cup \xbf)
$}
\STATE{$\Cbin{j} \gets \Cbin{j} \cup \xbf^{\star}$}
\STATE{$\C{j}{} \gets \C{j}{} \setminus \xbf^{\star}$}
\ENDWHILE
\ENDFOR
\caption{$split$ procedure \label{algo-split}}
\end{algorithmic}
\textbf{Output:} a partition $\hat{\Cbold} = split(\Cbold) = (\C{1}{},\dots, \C{\ell}{}, \Cbin{j_{1}},\dots,\Cbin{j_{\ellbin}})$
\end{algorithm}

\subsection{\textit{merge} phase}

The $merge$ procedure goes back to $\ell$ clusters by combining some of the $\ell + \ellbin$ clusters together. 
The approach here simply consists in testing all the possible mergings to go from $\ell + \ellbin$ to $\ell$ groups, and in keeping the one with the lowest quantization error. 
Algorithm~\ref{algo-merge} details the procedure. The set of all the perturbations tried, $G(\Cbold)$, is built in the algorithm and is the union of all the $\doublehat \Cbold$. As $\mathscr{P}$ is the set of all partitions of $1,\dots,\ell + \ellbin$ into $\ell$ groups, the number of possible mergings, which is the cardinal of $\mathscr{P}$, is equal to the Stirling number of the second kind $S(\ell + \ellbin,\ell)$ (\cite{Chan}). 
This number is reasonable if $\ellbin = 2$. For instance, $S(4,2) = 7$, $S(5,3) = 25$, $S(6,4) = 65$, $S(7,5) = 140$. 
Thus, we select that value of $\ellbin$ for the applications. 

\begin{algorithm}
\textbf{Input:} Sample $(\xbf_{i})_{i=1}^{n}$, a partition $\hat{\boldsymbol{C}} = (\hat{C}_{1},\dots, \hat{C}_{\ell + \ellbin})$
\begin{algorithmic}
\STATE $\Cbold^{\text{best}} \gets \emptyset$
\STATE $\mathcal{E}^{\star} \gets +\infty$
\STATE $\mathscr{P} = \{\parti$ \,: \; partition of ${1,\dots,\ell+\ellbin}$ in $\ell$ groups\}
\FOR $\parti \in \mathscr{P}$
\STATE $(G_{1},\dots, G_{\ell}) \gets \parti$ 
\FOR $j \in 1:\ell$
\STATE $\doublehat{C}_{j} = \bigcup_{k \in G_j} \hat{C}_{k}$
\ENDFOR 
\STATE $\doublehat{\Cbold} \gets (\doublehat{C}_{1},\dots,\doublehat{C}_{\ell})$
\STATE $\mathcal{E} \gets \errquant{p}(\doublehat{\Cbold})$
\IF{$\mathcal{E} <\mathcal{E}^{\star}$}
\STATE{ $\Cbold^{\text{best}} = (\C{1}{\text{best}}, \dots, \C{\ell}{\text{best}}) \gets \doublehat{\Cbold}$}
\STATE{ $\mathcal{E}^{\star} \gets \mathcal{E}$}
\ENDIF
\ENDFOR
\caption{$merge$ procedure \label{algo-merge}}
\end{algorithmic}
\textbf{Output:}   $\Cbold^{\text{best}} = (\C{1}{\text{best}}, \dots, \C{\ell}{\text{best}}) \gets \doublehat{\Cbold}$
\end{algorithm}

It is important to note that the clustering before splitting, $\Cbold$, is described by one of the partitions of $\mathscr{P}$, that is $\Cbold \in G(\Cbold)$.
Therefore, the best possible merge can return to the clustering before the perturbation step, which guarantees that $\perturb$ does not increase the quantization error (Proposition~\ref{prop_perturb}).

\subsection{Perturbation intensity}

In our implementation, the clustering perturbation intensity is set to decrease with time. 
Looking at the algorithm as a minimizer of the quantization error, this means that the search will be more exploratory at the beginning than at the end, as it is customary in stochastic, global, optimization methods such as simulated annealing.
The perturbation intensity is controlled by $\ellbin$ and $\nbin$. $\ellbin$ is set equal to 2 to keep the computational complexity of $merge$ low enough. $\nbin$ decreases with an a priori schedule made of 3 epochs where $\nbin=0.4$ then $0.2$ and $0.1$. 
These values were found by trial and error.
Within each epoch, several iterations of ($\rtoc$,$\perturb$,$\ctor$) are performed. 
Before explaining the stopping criterion, we need to describe the last step, $\ctor$.

\section{An implementation of the Augmented Quantization algorithm}
\label{detailedalgo}

The implemented version of the Augmented Quantization algorithm is detailed in Algorithm~\ref{detailed_augmented_algo} hereafter. It is based on a repetition of epochs during where the perturbation intensity (controlled by the relative size of the cluster bins, $\nbin$). An epoch is made of cycles of the $\rtoc$, $\perturb$ and $\ctor$ steps until convergence of the representatives or a maximum number of cycles. 
Between epochs, $\nbin$ is decresed.
An empirical tuning of the algorithm's parameters led to the following values: $\nbin$ decreases according to list\_$\nbin$ = $[0.4,,0.2,0.1]$.
A stopping criterion for each epoch is implemented in the form of either a minimal change in the estimated mixture, computed with the $p-$Wassertstein distance, or a maximal number of iterations. This maximal number of iterations is set to $\mathrm{it}_{\mathrm{max}} = 10$ and the convergence threshold in terms of distance between successive mixtures is MinDistance = $2m\times 10^{-3},$ with $m$ the dimension of the inputs space.

\begin{algorithm}
\textbf{Input:} a sample $(\xbf_{i})_{i=1}^{n}$, $\Rbold = (\R{1}{},\dots, \R{\ell}{}) \in \mathcal{R}^{\ell}$, list\_$\nbin$, $\mathrm{it}_{\mathrm{max}}$, minDistance
\begin{algorithmic}
\STATE{$J$ random variable with $P(J=j)=\frac{1}{\ell}$}
\STATE{$(\Rbold^{\star}, \Cbold^{\star}, \mathcal{E}^{\star}) \gets (\emptyset, \emptyset, +\infty)$}
\FOR{$\nbin \in$ list\_$\nbin$}
\STATE{dist $\gets +\infty$}
\STATE{it $\gets 1$}
\WHILE{dist $>$ minDistance AND it $\leq$ $\mathrm{it}_{\mathrm{max}}$}
\STATE{$\Rbold^{\mathrm{old}}_{J_\mathrm{old}} \gets \Rbold_J$}
\STATE{$\Cbold \gets \rtoc(\Rbold,J)$}
\STATE{$\Cbold \gets \perturb(\Cbold), \nbin)$}
\STATE {$\Rbold \gets \ctor(\Cbold)$}
\STATE{$J$ r.v. with $P(J=j) = \frac{\car\left(\C{j}{}\right)}{n},\: j \in \mathcal{J}$}
\IF{$\errquant{p}(\Cbold,\Rbold)<\mathcal{E}^\star$}
   $\mathcal{E}^\star \gets \mathcal{E} ~,~ \Cbold^{\star} \gets \Cbold ~,~ \Rbold^{\star} \gets \Rbold ~,~ J^{\star} \gets J $
\ENDIF
\STATE{it $\gets$ it $+1$}
\STATE{dist $\gets W_p(\Rbold^{\mathrm{old}}_{J_\mathrm{old}},\Rbold_J$)} 
\ENDWHILE
\ENDFOR
\end{algorithmic}
\textbf{Output:} $\R{J^{\star}}{\star}$
\caption{Detailed Augmented quantization algorithm}
\label{detailed_augmented_algo}
\end{algorithm}

\end{document}